\documentclass{statsoc}

\usepackage[a4paper]{geometry}
\usepackage{graphicx}
\usepackage[textwidth=8em,textsize=small]{todonotes}
\usepackage{amsmath}
\usepackage{amssymb}

\usepackage{algorithm}
\usepackage{algorithmic}
\usepackage{array} % \newcolumntype

\usepackage{natbib}
\usepackage[caption=false]{subfig}
\newtheorem{theorem}{Theorem}%[section]

\newtheorem{lemma}{Lemma}

\newtheorem{proposition}{Proposition}
\usepackage{thm-restate}

\newcolumntype{C}[1]{>{\centering\let\newline\\\arraybackslash\hspace{0pt}}m{#1}}

\usepackage{xcolor}
\usepackage{comment}

\newcommand\independent{\protect\mathpalette{\protect\independenT}{\perp}}
\def\independenT#1#2{\mathrel{\rlap{$#1#2$}\mkern2mu{#1#2}}}

\title[Uniformly Ergodic Block Sampling for SEM]{Exact Inference for Stochastic Epidemic Models via Uniformly Ergodic Block Sampling}
\author[Rapha\"{e}l Morsomme {\it et al.}]{Rapha\"{e}l Morsomme}
\address{Department of Statistical Science, Duke University, Durham, U.S.A.}
\email{raphael.morsomme@duke.edu}
\author[Morsomme and Xu]{Jason Xu}
\address{Department of Statistical Science, Duke University, Durham, U.S.A.}
\email{jason.q.xu@duke.edu}

\begin{document}
\begin{abstract}
    Stochastic epidemic models provide an interpretable probabilistic description of the spread of a disease through a population. Yet, fitting these models to partially observed data is a notoriously difficult task due to intractability of the likelihood for many classical models. To remedy this issue, this article introduces a novel data-augmented MCMC algorithm for  exact Bayesian inference under the stochastic SIR model, given only discretely observed counts of infection. In a Metropolis-Hastings step, the latent data are jointly proposed from a surrogate process carefully designed to closely resemble the SIR model, from which we can efficiently generate epidemics consistent with the observed data. This yields a method that explores the high-dimensional latent space efficiently, and scales to outbreaks with hundreds of thousands of individuals. We show that the Markov chain underlying the algorithm is uniformly ergodic, and validate its performance via thorough simulation experiments and a case study on the 2013-2015 outbreak of Ebola Haemorrhagic Fever in Western Africa.
\end{abstract}


\section{Introduction}  \label{sec:intro}

The efficient control of a disease outbreak requires an understanding of the mechanisms underlying its spread among individuals. Mechanistic compartmental models, which describe the transition of individuals between various disease states, have a long mathematical modeling tradition in epidemiology dating back to \cite{Kermack.1927}. Due to their interpretability, these models are commonly used to describe the dynamics of an outbreak and typically serve as the main source of information for predicting its course and identifying interventions that could be effective \citep{Anderson.1992}. Originally, deterministic versions of the models were employed by mathematicians and epidemiologists. These models are simple to analyze, but fail to capture the inherent randomness characterizing the spread of a disease. For instance, they cannot be used to estimate the probability of a large-scale outbreak or its expected duration, and do not allow for uncertainty quantification when used within inferential procedures. Stochastic epidemic models (SEM), on the other hand, incorporate the random nature of infections and recoveries and therefore provide more realistic descriptions of the spread of a disease, and in turn more reliable inference from observed data.

Conducting inference on general SEMs is, however, a notably difficult task. Challenges stem from the fact that the observed data typically provide incomplete information on a process that evolves continuously through time, making the likelihood of the model intractable. The marginal likelihood of such partially observed data becomes a computational bottleneck, as it requires a large integration step that accounts for all possible trajectories of the missing epidemic process.
Direct computation of this likelihood requires the transition probabilities between observation times for which no closed form is available. Moreover, given the large size of the transition matrix, classical matrix exponentiation is intractable. \cite{Ho.2018} recently developed a numerical method to compute these transition probabilities, but their method only applies to Markovian SEMs, and high computational costs limit their approach to moderate sized outbreaks.
To bypass the intractable marginal likelihood of SEMs, \cite{King.2015} develop particle filters, which are popular among practitioners, and \cite{McKinley.2018} use the approximate Bayesian computation (ABC) framework. However, these forward simulation methods are computationally intensive, and can become degenerate when the model is misspecified.

As an alternative to direct marginalization and forward simulation, one can account for the missing data via sampling. MCMC algorithms based on data augmentation (DA) can facilitate Bayesian inference of partially observed SEMs by exploring configurations of the missing data through latent variables.
A first drawback of existing DA-MCMC algorithms is their limitations to observed data that consist in either the exact removal times \citep{Gibson.1998, ONeill.1999} or in discretely observed prevalence data \citep{Fintzi.2017}. This severely limits their applicability since such detailed data are only observed in animal populations or in small outbreaks. Instead, public health surveillance systems often report incidence data such as weekly counts of infections or deaths.  \cite{Lekone.2006} designed a DA-MCMC for fitting discrete-time SEMs to incidence data, but to our knowledge, no such method exists for continuous-time SEMs that can efficiently explore the latent space. Past attempts employ single-site samplers to update the high-dimensional latent data in which a single element is modified per iteration. Such samplers are known to mix slowly when the latent variables are correlated, as is the case in epidemic models, resulting in highly correlated MCMC draws. This confines these methods to small outbreaks for which the Markov chain can be expected to mix sufficiently in a reasonable amount of time. In contrast, block samplers update the entire latent data jointly, updating correlated variables simultaneously. When they can be effectively implemented, block samplers allow the chain to move more quickly through the high-dimensional latent space, often considerably improving mixing. In concurrent work, \cite{Wang.2022} propose a block sampler for SEM, but their approach assumes that the infection times or the removal times are observed exactly, which is rarely the case in practice.

This article introduces a novel DA-MCMC algorithm for fitting continuous-time SEMs to incidence data. Using a novel idea that decouples non-linearities in each observation  period, our method admits an efficient block sampler that scales well to large outbreaks.
The algorithm updates the event times comprising the latent data with a block Metropolis-Hastings (M-H) step, jointly proposing the infection and removal times from a surrogate stochastic process whose dynamics closely resemble those of the target SEM. In particular, doing so allows us to efficiently generate an epidemic that is consistent with the observed data.
The idea of decoupling via more tractable processes applies generally; in our setting, we employ a multitype branching process that faithfully approximation the target SEM. A large proportion of the latent event times can be updated per iteration with high acceptance rate. Not only does the chain explore the high-dimensional latent space efficiently, but we derive theoretical guarantees not available for past approaches in this class, including uniform ergodicity of the sampler.
Our approach is fully Bayesian, allowing representation of parameter uncertainty through posterior distributions, and the proposed DA-MCMC targets the exact posterior distribution of model parameters under the target SEM.
	
The remainder of the article is structured as follows. Section \ref{sec:set} provides background information on the inference task. It introduces the SEM that we use to illustrate the proposed DA-MCMC and explains why conducting inference from partially observed epidemic data is a difficult task. Previous works addressing this problem are also presented. Section \ref{sec:pds} introduces the DA-MCMC algorithm and describes the surrogate process from which the latent data are generated in the M-H step. Important features of this surrogate process is discussed in Section \ref{sec:imp}, and the DA-MCMC is shown to be uniformly ergodic in Section \ref{sec:uni}. In Section \ref{sec:sim}, we examine the performance of our algorithm on simulated data and then turn to an analysis of the 2013-2015 Ebola outbreak in Gu\'eck\'edou, Guinea, in Section \ref{sec:ebo}. Finally, Section \ref{sec:dis} discusses the findings and concludes the article.

\section{Background and Prior Work}  \label{sec:set}

\subsection{The Non-Markovian Stochastic SIR Model}  \label{sec:sir}

%\jx{rephrase beginning of this section to emphasize generality of our method}

Our point of departure is to consider the stochastic SIR model \citep{Bailey.1975}, a compartmental model which offers a parsimonious and interpretable representation of the spread of a contagious disease in a population.
While our methodology readily applies to more complex SEMs such as the SEIR model and related processes, in this article we focus our attention on SIR model, but later allow for non-Markovian processes under which infectious periods follow an arbitrary distribution  \citep{Streftaris.2002, Lloyd.2001}.
Under this model, individuals transition through three states or compartments: susceptible (S), infectious (I) and removed (R). The only possible moves are from S to I (infections) and from I to R (removals). A susceptible individual becomes infected through contact with an infectious individual. Once infected, she is immediately infectious and remains so for some period of time after which she is removed from the process without the possibility of reinfection.

Assuming a closed population of $n$ individuals, the stochastic SIR model consists of a continuous-time vector-valued process
\begin{equation}
	\label{eq:X}
	\mathbf{X} = \left\lbrace \mathbf{X}(t) = \left(X_1(t), \dots, X_n(t)\right), t>0\right\rbrace \in \chi_{\mathbf{X}}
\end{equation}
where the agent-level subprocess
\begin{equation*}
X_j(t) = 
\begin{cases}
	s, & t \in [0, \tau^I_j] \\
	i, & t \in (\tau^I_j, \tau^R_j] \\
	r, & t \in (\tau^R_j, \infty)
\end{cases}
,\quad i = 1, \dots, n
\end{equation*}
denotes the compartment of individual $j$ at time $t$ with $\tau^I_j$ and $\tau^R_j$ respectively the infection and removal times of individual $j$. If individual $j$ is never infected, we set $\tau^I_j = \tau^R_j = \infty$ and $X_j(t) = s$ for $t \in [0, \infty)$. As removed individuals do not contribute to the pandemic, we can ignore the individuals removed at time $0$ and write $n = S(0) + I(0)$ ($S(t)$ and $I(t)$ are the sizes of the susceptible and infective populations at time $t$). Here the space $\chi_{\mathbf{X}}$ denotes the set of possible trajectories for $\mathbf{X}$---that is, the set of trajectories in which no infection occurs after the infectious compartment is depleted:
\begin{equation*}
	\chi_{\mathbf{X}} = \{\mathbf{X}:\mathbf{X}(t) \in \{s,r\}^n \Rightarrow \mathbf{X}(t+u) = \mathbf{X}(t), \forall u>0 \}.
\end{equation*}

We assume a homogeneously mixing population of exchangeable individuals in which contacts between each pair of individuals follow independent Poisson processes with positive rate $\beta \in \mathbb{R}^+$. Since at time $t$ there are $I(t)$ infective individuals, the effective infection rate of each susceptible individual is $\beta I(t)$. The infectious periods are identically and independently distributed (i.i.d.) random variables following an arbitrary distribution $\mathcal{F}$ with unknown parameter $\lambda \in \chi_\lambda$. While practitioners often model infectious periods using the exponential distribution in order to obtain a Markov process, \cite{Lloyd.2001} suggests to use a less dispersed distribution on the ground of realism. In Section \ref{sec:per}, we assume that infectious periods follow a Weibull distribution, a flexible and computationally convenient family of distributions of which the exponential distribution is a special case, with unknown scale parameter $\lambda$ and known shape parameter, in which case $\chi_\lambda = \mathbb{R}^+$. When the process \eqref{eq:X} is observed until time $T$, the likelihood is \citep{Streftaris.2002}
\begin{align}
	\label{eq:cdl}
	L(\theta; \mathbf{X})
	& = \prod_{j \in \mathcal{I}} \beta I(\tau^I_j) \exp\left\lbrace - \int_{0}^{T}\beta S(t)I(t)dt \right\rbrace \prod_{k \in \mathcal{R}} f(\tau^R_k - \tau^I_k;\lambda) \prod_{l \in \mathcal{R}^c} \bar{F}(T - \tau^I_k;\lambda) %\nonumber \\
%	& = \beta^{n_I} \lambda^{n_R} a^{n_R} \prod_{j \in \mathcal{I}} I(\tau^I_j) \exp\left\lbrace - \beta \int_{0}^{T} S(t)I(t)dt \right\rbrace 
%	\prod_{k \in \mathcal{R}} (\tau^R_k - \tau^I_k)^{a-1} \nonumber \\
%	\label{eq:cdl}
%	& \qquad \exp\left\lbrace - \lambda \left[ \sum_{k \in \mathcal{R}} (\tau^R_k - \tau^I_k)^a + \sum_{l \in \mathcal{R}^c} (T - \tau^I_l)^a\right] \right\rbrace.
\end{align}
To establish notation, $\theta = (\beta, \lambda) \in \chi_\theta$ are the unknown model parameters and $\chi_\theta = \mathbb{R}^+ \times \chi_\lambda$ is the parameter space, the index sets $\mathcal{I} = \{j: \tau^I_j \in (0, T]\}$,  $\mathcal{R} = \{j: \tau^R_j \in (0, T]\}$ and $\mathcal{R}^c = \mathcal{I} \setminus \mathcal{R}$ respectively denote the individuals that are infected, removed, and infected but not removed during the observation interval $(0, T]$, $n_I$ and $n_R$ are the numbers of observed infections and removals, and $\bar{F}(.)$ is the survival function of $\mathcal{F}$. Since the functions $I(t)$ and $S(t)$ are constant between event times, the integral corresponds to a finite sum which is straightforward to compute.

\subsection{Inference with Complete and Incomplete Data}  \label{sec:icd}
%Outline: complete data likelihood; MLE; gamma conjugacy;
% observed data; intractable posterior; DAMCMC; parameters update; latent space update

When the process \eqref{eq:X} is completely observed until time $T$, conducting inference for $\beta$ is straightforward. The likelihood of $\beta$ belongs to the exponential family and Bayesian inference is facilitated by the conjugacy of \eqref{eq:cdl} with the gamma distribution. If we let
\begin{equation}  \label{eq:pri_beta}
	\beta \sim \textsf{Ga}(a_{\beta}, b_{\beta})%, \qquad \lambda \sim \textsf{Ga}(a_{\lambda}, b_{\lambda}),
\end{equation}
independently of $\lambda$, where $\textsf{Ga}(a,b)$ denotes the gamma distribution with mean $a/b$ and variance $a/b^2$, then the full conditional distribution of $\beta$ is
\begin{equation}  \label{eq:posterior_beta}
	\beta | \mathbf{X}, \lambda \sim \textsf{Ga}\left( a_{\beta} + n_I , b_{\beta} + \int_{0}^{T} I(t)S(t) dt\right)
\end{equation}
and does not depend on $\lambda$. 

In Section \ref{sec:per}, we let infection periods follow a Weibull distribution. Appendix $\ref{app:wei}$ shows that this choice of distribution leads to straightforward inference for the Weibull scale parameter $\lambda$ given a fixed shape parameter $a$. In particular, if
\begin{equation}  \label{eq:pri_lambda}
	\lambda \sim \textsf{Ga}(a_{\lambda}, b_{\lambda})
\end{equation}
independently of $\beta$, then the full conditional of $\lambda$ is
\begin{equation}  \label{eq:posterior_lambda}
	\lambda | \mathbf{X}, \beta \sim \textsf{Ga}\left( a_{\lambda} + n_R, b_{\lambda} + \sum_{k \in \mathcal{R}} (\tau^R_k - \tau^I_k)^a + \sum_{l \in \mathcal{R}^c} (T - \tau^I_l)^a \right)
\end{equation}
and does not depend on $\beta$. In the complete case setting, this choice of $\mathcal{F}$ therefore leads to a straightforward exploration of $\pi(\theta|\mathbf{X})$ via Monte Carlo draws from the two independent distributions \eqref{eq:posterior_beta} and \eqref{eq:posterior_lambda}.

In practice, however, inference is complicated by the fact that we rarely observe the epidemic process \eqref{eq:X} in such detail. When some event times are unobserved, we cannot evaluate the likelihood \eqref{eq:cdl}.
%Various types of partially observed data typify real data and have been considered in the literature. These include the removal times \citep{Gibson.1998, ONeill.1999} or the number of infectious individuals at discrete points in time \citep{Fintzi.2017}. 
% For instance, the former arises in animal experiments in which positive cases are immediately isolated from the rest of the population and no longer contribute to disease spread. In such cases, exact times of removals are known, but this information is unavailable in a typical observational study.
In this article, we focus our attention on discretely observed incidence data, e.g.\ weekly infection counts.
Given time intervals $\mathcal{T} = \{\mathcal{T}_1, \dots, \mathcal{T}_K:\mathcal{T}_k = (t_{k-1}, t_k]\}$, such that $t_0=0$ and $t_K = T$, the observed data consist of the $K$-dimensional vector $\mathbf{Y} = (I_1,\dots,I_K)$ where $I_k := \#\{\tau^I_j \in \mathcal{T}_k\}$, ($k=1,\dots,K$) is the number of infections in the $k^{\text{th}}$ interval. In words, we do not know the exact infection times but only the interval in which they took place.

This form of observed data is often reported by public health surveillance systems \citep{Fintzi.2020}.
As motivation, consider the $2013$-$2015$ outbreak of the Ebola virus in Western Africa which resulted in more than 11,000 deaths, %, the largest outbreak since the discovery of the virus in 1976.
where the available data consist of the weekly numbers of positive tests in each province of Guinea, Liberia and Sierra Leone, the three countries where the vast majority of fatalities occurred. Individuals infected by the Ebola virus becomes infectious only once the symptoms start \citep{Coltart.2017}. Moreover, unlike contagious diseases for which the isolation of infected individuals effectively prevents them from infecting other individuals, the 2013-2015 outbreak was characterized by a large number of infections that occurred after individuals were tested positive for the Ebola virus. \cite{Coltart.2017} report that numerous infections occurred at the hospitals where infected individuals received treatment as well as at the funerals of people deceased from the virus\footnote{In most regions impacted by the virus, touching the body of the deceased during funerals is a tradition. Since the Ebola virus is transmitted via sweat, numerous infections occurred at funerals.}. 
Finally, the severity of the symptoms suggests that the vast majority of the infections were reported.
We therefore model the number of positive tests reported by the WHO in a given week as the number of infections in the SIR model. Although the SEIR process is a more adequate model for the Ebola virus which is characterized by a latent period, in this article we opt for the simpler SIR process for illustration and note that our DA-MCMC algorithm can easily be extended to the SEIR.

In a Bayesian framework, the posterior distribution of the parameters given the observed data is formally related to the complete data likelihood \eqref{eq:cdl} via integration:
\begin{equation}
	\label{eq:pdl}
	\pi(\theta|\mathbf{Y}) 
	\propto \pi(\theta) L(\theta; \mathbf{Y}) = \pi(\theta) \int_{\chi_\mathbf{x}} f(\mathbf{Y}|\mathbf{x}) L(\theta; \mathbf{x}) d\mathbf{x}
	%\mathbf{1}\{\Y \overset{\Delta}{=} \mathbf{x}\}
\end{equation}
where $\pi(\theta)$ is the prior distribution on $\theta$ and $f(\mathbf{Y}|\mathbf{x})$ is the likelihood of the observed data conditionally on the epidemic process. The partial data likelihood $L(\theta; \mathbf{Y})$ consists of a high-dimensional integral over all epidemic paths resulting in the observed data. This marginalization step has no known analytical solution and presents computational challenges even for a population of moderate size \citep{Ho.2018}.

\subsection{Prior Work}  \label{sec:pre}

Several approaches for conducting inference on partially observed stochastic epidemic models have been proposed. Approximation methods are based on a simpler process that approximates the model's dynamics and whose likelihood in the presence of partially observed data is tractable. Popular approximations include chain binomial models \citep{Greenwood.1931, Abbey.1952}, diffusion processes \citep{Cauchemez.2008, Fintzi.2020}. While these approximations bypass the intractability of the original likelihood \eqref{eq:pdl}, the assumptions on which these methods based are questionable in typical applied settings. The discrete-time approximations used in chain binomial models are questionable when the time scale of the observation process is long, and diffusion approximations of the epidemic process are only valid for very large epidemics.

Two families of sampling-based methods have been proposed to directly work with the partial likelihood instead of an approximation thereof: model-based forward simulation and data-augmented MCMC (DA-MCMC).
Particle filtering is an example of the former category that is popular among practitioners \citep{King.2015}. Its plug-and-play feature makes it applicable to a wide variety of models. However, model-based forward simulation suffers from two drawbacks: simulating data from a model as complex as the SIR that is consistent with the observed data is prohibitively slow, and these methods can fail to converge when the model does not fit the data well. The approximate Bayesian computation (ABC) framework offers a solution to the latter problem \citep{McKinley.2018}, but its inference is based on an approximation of the model's likelihood. As a result, the inference can be biased, and it is difficult to evaluate the degree of approximation involved. %quantify how closely the approximate posterior resembles the original target distribution.

DA-MCMC algorithms treat the unobserved epidemic events as nuisance parameters. Under this approach, a MCMC algorithm is used to explore the joint distribution of model parameters and unobserved event times. The resulting sample space is high-dimensional, which complicates inference. \cite{Gibson.1998} and \cite{ONeill.1999} employed the reversible-jump MCMC framework \citep{Green.1995} to explore models with different numbers of unobserved events. 
These authors explore the latent space by uniformly inserting, deleting or moving a single latent variable per iteration.
More recently, \cite{Fintzi.2017} designed a Gibbs-like sampler that updates one latent variable at a time from a distribution that closely resembles its full conditional distribution, \cite{Bu.2020} constructed a single-site Gibbs sampler for epidemics over networks when the exact infection times are assumed to be known, and \cite{Touloupou.2020} constructed a Gibbs sampler that updates the trajectory of a single individual per iteration in the discrete-time setting.
	
These DA-MCMC methods are prone to poor mixing in the presence of large epidemics. Since they update at most a very small number of latent variables at a time, the resulting Markov chain is sticky: it makes very small jumps in the latent space and therefore requires a large number of iterations to explore it completely.
Some tweaks, such as updating slightly more event times per iteration in \cite{Pooley.2015} or non-centered parameterizations \citep{Neal.2005} may ameliorate the issue, but the effective gains are modest. The resulting chains continue to suffer from very high auto-correlation, and do not possess adequate mixing properties in populations over a few hundred individuals.
By a slight abuse of language, we will refer to these samplers for the latent data as \textit{single}-site samplers to contrast them to the block sampler proposed in this article.

\section{Exact Inference via Data-Augmentation}  \label{sec:con}

We adopt a DA-MCMC approach that bridges the challenging partially observed setting to the tractable complete data likelihood by way of latent variables. The M-H block sampler that we propose for the latent data hinges on a carefully designed proposal process that faithfully approximates the SEM, and from which we can trivially generate an epidemic consistent with the observed incidence counts.

As we saw in Section \ref{sec:icd}, the complete data likelihood \eqref{eq:cdl} is amenable to computation. This suggests augmenting the observed data $\mathbf{Y}$ with latent data $\mathbf{Z}$ such that the likelihood $L(\theta; (\mathbf{Y}, \mathbf{Z}))$ has the closed form \eqref{eq:cdl}, and constructing a MCMC algorithm that iterates between updates of the parameters and of the latent data to generate an ergodic Markov chain $\{(\theta^{(m)}, \mathbf{Z}^{(m)})\}_{m=0}^M$ whose limiting distribution is the joint posterior $\pi(\theta, \mathbf{Z}|\mathbf{Y})$. Given MCMC draws $\{(\theta^{(m)}, \mathbf{Z}^{(m)})\}_{m=0}^M$, the sequence $\{\theta^{(m)}\}_{m=0}^M$ provides an approximation of the marginal posterior of interest $\pi(\theta|\mathbf{Y})$.

For the stochastic SIR, the latent data $\mathbf{Z} = \left\lbrace (z^I_j, z^R_j)\right\rbrace_{j=1}^{n}$ consist of the infection and removal times of each individual which are both unobserved. If individual $j$ is not infected (removed) before $T$, we write $Z^I_j=\infty$ ($Z^R_j=\infty$). Note that latent data $\mathbf{Z}$ contains the same information as the complete data latent data $\mathbf{X}$.
%	$$z^I_j \begin{cases}
%		\in [0, T) & \text{if individual } j \text{ is infected before } T \\
%		= \infty & \text{if individual } j \text{ is not infected before } T \\
%	\end{cases}$$
%	and
%	$$z^R_j \begin{cases}
%		\in (z^I_j, T] & \text{if individual } j \text{ is removed before } T \\
%		= \infty & \text{if } z^I_j = \infty \text{ or individual } j \text{ is not removed before } T. \\
%	\end{cases}$$
Since $\mathbf{Y} = (I_1,\dots,I_K)$ is a deterministic function of the infection events, the function $f(\mathbf{Y}|\mathbf{X})$ in \eqref{eq:pdl} is equal to $\delta_{\mathbf{Y}}(\mathbf{X})$, an indicator function which is $1$ if the epidemic process $\mathbf{X}$ results in the observed infection counts $\mathbf{Y}$ and $0$ otherwise. The joint posterior distribution is therefore $\pi(\theta, \mathbf{Z}|\mathbf{Y}) \propto \delta_{\mathbf{Y}}(\mathbf{Z}) L(\theta; \mathbf{Z}) \pi(\theta)$, where $L(\theta; \mathbf{Z})$ is the complete data likelihood \eqref{eq:cdl}.

We construct a DA-MCMC algorithm that iteratively updates $\theta$ and $\mathbf{Z}$ conditionally on each other and $\mathbf{Y}$, based on $\pi(\theta|\mathbf{Z}, \mathbf{Y})$ and $\pi(\mathbf{Z}|\theta, \mathbf{Y})$. A one-step transition from $\mathbf{x}_1 = (\theta_1, \mathbf{z}_1)$ to $\mathbf{x}_2 = (\theta_2, \mathbf{z}_2)$ therefore looks like
$$
\mathbf{x}_1 = (\theta_1, \mathbf{z}_1) \rightarrow (\theta_2, \mathbf{z}_1) \rightarrow (\theta_2, \mathbf{z}_2) = \mathbf{x}_2.
$$
The parameter updating step is straightforward. In the case of Weibull-distributed infection periods, sampling from $\pi(\theta|\mathbf{Z}, \mathbf{Y})$ amounts to a Gibbs step in which we sample from the two independent full conditional distributions \eqref{eq:posterior_beta} and \eqref{eq:posterior_lambda}, treating the latent data as the complete data.
In contrast, updating the latent data is difficult. Though unrestricted forward simulation of SEMs is straightforward \citep{Gillespie.1977}, drawing trajectories conditionally on the incidence counts $\mathbf{Y}$ amounts to \textit{conditional} simulation of a stochastic process, a notoriously difficult task. Here the methods proposed by \cite{Hobolth.2009} are not applicable since the process is not necessarily Markovian, and given $\mathbf{Y}$ the end-point of the process are not known.
Instead, to update the latent data we employ a M-H block sampler in which a complete new configuration of $\mathbf{Z}$ consistent with $\mathbf{Y}$ is generated from a surrogate process which we present in Section \ref{sec:pds}. Given some current $\theta$ and $\mathbf{Z}$, the M-H step proceeds by simulating a candidate $\mathbf{Z}^\star$ from the distribution $Q(.|\theta, \mathbf{Y})$ with density $q(.|\theta, \mathbf{Y})$ corresponding to the surrogate process. Then $\mathbf{Z}^\star$ is accepted with probability
\begin{equation}  \label{eq:alpha}
	\alpha\left( \left( \theta, \mathbf{Z}\right) , \left( \theta, \mathbf{Z}^\star\right) \right) =	\min\left\lbrace 1, \dfrac{L\left( \theta; \mathbf{Z}^\star\right) q\left( \mathbf{Z}|\theta\right)}{L\left( \theta;\mathbf{Z}\right) q\left( \mathbf{Z}^\star|\theta\right)}\right\rbrace,
\end{equation}
and otherwise the current $\mathbf{Z}$ is retained. Note that it is not necessary to compute $\delta_{\mathbf{Y}}(\mathbf{Z})$ in the M-H acceptance ratio \eqref{eq:alpha} since the proposal distribution $Q$ is designed to ensure that $\mathbf{Z}^\star$ is consistent with $\mathbf{Y}$.

%\begin{algorithm}
%	\caption{Data-Augmented MCMC Sampler}
%	\label{alg:DA-MCMC}
%	\begin{algorithmic}
%		\REQUIRE $\theta^{(0)}$
%		\RETURN $\{(\mathbf{z}^{(m)}, \theta^{(m)})\}_{m=0}^N$ 
%		\STATE $\mathbf{z}^{(0)} \sim Q(.|\theta^{(0)})$ (generate the initial latent data from the PD-SIR process)
%		\FOR {$j = 1, \dots, N$}
%		\STATE $\beta^{(j)}|\mathbf{z}^{(j - 1)} \sim Ga\left( a_{\beta} + n_I^{(j - 1)}, b_{\beta} + \int_{0}^{T} I^{(j - 1)}(t)S^{(j - 1)}(t)dt\right)$ (Gibbs update)
%		\STATE $\lambda^{(j)}|\mathbf{z}^{(j-1)} \sim Ga\left( a_{\lambda} + n_R^{(j - 1)}, b_{\lambda} + \int_{0}^{T} I^{(j - 1)}(t) dt\right)$ (Gibbs update)
%		\STATE $\theta^{(j)} \leftarrow (\beta^{(j)}, \lambda^{(j)})$
%		\STATE $\mathbf{z}^\star \sim Q(.|\theta^{(j)})$ (generate latent data from the PD-SIR process)
%		\STATE $\alpha = \min\left\lbrace 1, \dfrac{L(\theta^{(j)}; \mathbf{z}^\star)q(\mathbf{z}^{(j - 1)}|\theta^{(j)})}{L(\theta^{(j)}; \mathbf{z}^{(j - 1)})q(\mathbf{z}^\star|\theta^{(j)})}\right\rbrace  $
%		\STATE $u \sim U(0,1)$
%		\IF{$u<\alpha$}
%		\STATE $\mathbf{z}^{(j)} \leftarrow \mathbf{z}^*$			
%		\ELSE
%		\STATE $\mathbf{z}^{(j)} \leftarrow \mathbf{z}^{(j - 1)}$
%		\ENDIF
%		\ENDFOR
%	\end{algorithmic}
%\end{algorithm}

\subsection{Block Sampler for the Latent Data}  \label{sec:pds}
% Outline: define PD-SIR; generate PD-SIR

The surrogate process used for proposing latent data in the M-H step, referred to as the \textit{piecewise decoupled SIR} process (PD-SIR), is a stochastic process whose dynamics closely resemble those of the SIR and which is designed for efficient simulation of epidemic trajectories consistent with $\mathbf{Y} = (I_1, \dots, I_K)$. 

Similarly to the SIR, the PD-SIR process corresponds to a compartmental model in which individuals move from the compartments $S$ to $I$ and from $I$ to $R$.
The removal dynamics are identical under both processes: infectious periods are i.i.d.\ draws from $\mathcal{F}$.

\begin{figure}
	\centering
	\includegraphics[width = .5\textwidth]{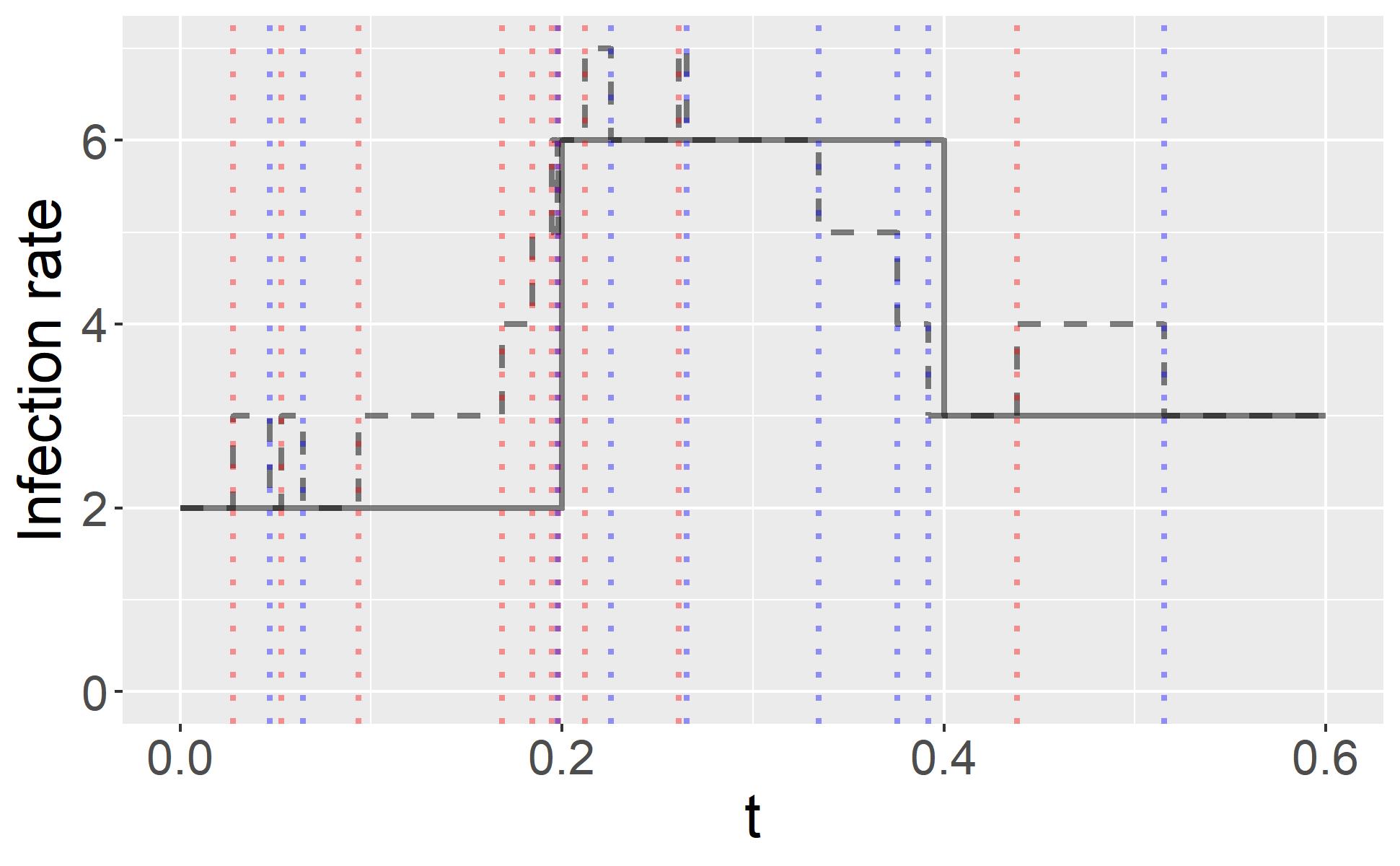}
	
	\caption{Infection rate of the SIR (dashed line) and PD-SIR (solid line) processes in a small population ($(S(0), I(0)) = (10,2)$) with $\beta=1$. The PD-SIR rate is reset at times $t_{0:2}=(0, 0.2, 0.4)$. The SIR rate varies after each infection (red) and removal (blue).}
	\label{fig:mu}
\end{figure}

The infection dynamics, however, differ slightly.
In the SIR process, the individual-level infection rate at time $t$ is $\mu(t) = \beta I(t)$.
The rate $\mu$ varies after every event since the value of $I$ changes after each infection or removal. This complicates the simulation of a SIR process consistent with the observed incidence counts.
In contrast, in the PD-SIR, the infection rate is kept constant over each time interval in $\mathcal{T} = \{\mathcal{T}_1, \dots, \mathcal{T}_K\}$,
$$
\tilde{\mu}(t) := \beta I(t_{k - 1}) = \mu_k, \quad t \in \mathcal{T}_k
$$
where $I(t_{k-1})$ is the number of infectious individuals at the beginning of the $k$th interval. As shown in Figure \ref{fig:mu}, $\tilde{\mu}$ and $I$ are decoupled in each interval, and the infection rate is reset at the start %or left endpoint
of the intervals. Over a single interval, the PD-SIR process is equivalent to a two-type branching process approximation of the SIR \citep{Ho.2018b}.

Under the PD-SIR, $S(t)$ follows a linear death process (LDP) during each interval $\mathcal{T}_k$ with death rate $\mu_k$, where infections in the PD-SIR corresponds to deaths in the LDP. Theorem \ref{theo:ldp}, whose proof is deferred to Appendix \ref{app:ldp}, shows that the event times of a LDP can easily be simulated conditionally on the number of events \citep{Neuts.1971}.
%The LDP is a counting process possessing the so-called order statistics property \citep{Neuts.1971}; that is, given the number of events registered on some interval, the event times are distributed as the order statistics of independent and identically distributed (i.i.d.) random variables following, in the case of the LDP, a exponential distribution truncated to that interval.
\begin{theorem}  \label{theo:ldp}
	Consider a linear death process with death rate $\mu$ and let $\tau_{1:N} \in (t_l, t_u]$ be the times of the $N$ deaths occurring between times $t_l$ and $t_u$. Then 
	$$\tau_j \,{\buildrel d \over =}\, X_{(j)}, \quad j = 1, \dots, N$$
	where $X_{(j)}$ is the $j^{\text{th}}$ order statistics of $N$ i.i.d.\ random variables following a truncated exponential distribution with rate $\mu$, lower bound $t_l$ and upper bound $t_u$.
\end{theorem}

We can use Theorem \ref{theo:ldp} to generate latent data $\mathbf{z} = \left\lbrace (z^I_i, z^R_i)\right\rbrace_{i=1}^{n}$ from the PD-SIR consistent with $\mathbf{Y}$. For each $k = 1, \dots, K$, we let $\mathcal{I}_k$ denote the indices of the $I_k$ individuals infected during interval $\mathcal{T}_k$ and realize the following two steps. First, following Theorem \ref{theo:ldp}, we generate the infection times $z^I_i$ from i.i.d.\ truncated exponential random variables with rate $\mu_k = \beta I(t_{k-1})$ bounded between $t_{k-1}$ and $t_k$,
$$
z^I_j \sim \textsf{TrunExp}(z^I_j; \mu_k, t_{k-1}, t_k), \quad j \in \mathcal{I}_k
$$
Note that $I(t_{k-1})$ only depends on past events and can therefore be computed given the PD-SIR process up to time $t_{k-1}$.
%Algorithm \ref{alg:PD-SIR} provides a simple recursion to compute this variable efficiently.
Moreover, since individuals in the PD-SIR are exchangeable, we do not need to compute the order statistics of the simulated values.  

Second, we generate the removal times $z^R_i$ of the same individuals from the exact removal dynamics of the SIR. To accomplish this, we independently sample the removal time of individual $j$ from the mixed distribution 
$$
z^R_j|z^I_j \sim (1 - p_j) \delta_{\infty}(z^R_j) + p_j \textsf{TrunF}(z^R_j - z^I_j; \lambda, 0, T - z^I_j), \quad j \in \mathcal{I}_k
$$
placing point mass $(1 - p_j)$ at $\infty$ and continuous mass on the interval $(z^I_j, T]$,
where $\delta_{\infty}$ corresponds to the Dirac distribution with mass $1$ on the element $\infty$, 
$$p_j = P(z^R_j \le T | z^I_j)$$
is the cumulative distribution function (CDF) of $\mathcal{F}$ and corresponds to the probability that individual $j$ is removed before $T$ given that she was infected at time $z^I_j$, and $\textsf{TrunF}(.; 0, T - z^I_j)$ denotes the truncated distribution $\mathcal{F}$ bounded between $0$ and $T - z^I_j$ for the infection period $(z^R_j - z^I_j)$.

Note that we do not assume that the outbreak is over: infected individuals are not necessarily removed during the observation period, in which case $z^R_j = \infty$. By assigning a value to $z^R_.$ for individuals removed after the end of the observation window, the dimension of the latent data remains constant across iterations. Unlike \cite{Gibson.1998} and \cite{ONeill.1999}, we therefore do not need to use reversible-jump MCMC to explore configurations of the latent data with different numbers of observed removals. % is this an example of \textit{saturation}?

By construction, this scheme, which is summarized in Algorithm \ref{alg:PD-SIR}, generates latent data from the PD-SIR process that are consistent with $\mathbf{Y} = (I_1, \dots, I_K)$. Its density is
\begin{alignat}{3}  \label{eq:q}
	q(\mathbf{z}|\theta) 
	& = && \prod_{k=1}^K \prod_{j\in\mathcal{I}_k} \textsf{TrunExp}(z^I_j; \mu_{k}, t_{k-1}, t_{k}) \nonumber \\
	& && \times \prod_{i=1}^{n} \left( 1 - p_i \right)^{\mathbf{1}(z^R_i = \infty)} \left( p_i \textsf{TrunF}(z^R_i - z^I_i; \theta, 0, T - z^I_i) \right)^{\mathbf{1}(z^R_i \le T)}  \nonumber \\
	& = && \prod_{k=1}^K \prod_{j\in\mathcal{I}_k} \textsf{TrunExp}(z^I_j; \mu_{k}, t_{k-1}, t_{k})
	\times \prod_{j \in \mathcal{R}^c} \bar{F}(T - z^I_k; \theta) \prod_{k \in \mathcal{R}}f(z^R_k - z^I_k; \theta)
\end{alignat}
where
$$
\textsf{TrunExp}(x; \mu, l, u) = \dfrac{\mu \exp\{-\mu x\}}{\exp\{-\mu l\} - \exp\{-\mu u\}}, \quad x \in (l, u)
$$
and
$$
\textsf{TrunF}(x; \theta, 0, u) = \dfrac{f(x; \theta)}{F(x; \theta)}, \quad x \in (0, u)
$$
respectively denote the density of a truncated exponential and the truncated distribution $\mathcal{F}$ with parameters as notated previously.

If we let $\mathcal{F}$ be the Weibull distribution, simulating from the PD-SIR process is straightforward as all the necessary random variables, such as the truncated exponential and truncated Weibull, can be generated via the inverse CDF method. Moreover, ensuring that the proposed latent data are consistent with the observed data is accomplished at no additional cost, making the PD-SIR scalable to large outbreaks.

\begin{algorithm}
	\caption{Generating a PD-SIR process conditionally on the observed data $\mathbf{Y}$}
	\label{alg:PD-SIR}
	\begin{algorithmic}
		\REQUIRE $\mathbf{Y} = (I_1, \dots, I_K), \theta = (\beta, \lambda)$, $I(0)$
%			\FOR {$j \in  \mathcal{I}_0$}
%			    \STATE $z^I_j \leftarrow 0$ (infection times)
%			    \STATE $p_j \leftarrow 1 - \exp\{-\lambda (T - 0)\}$
%			    \STATE $z^R_j \sim (1 - p_j) \delta_\infty + p_j \textsf{TrunWeib}(\lambda, a; 0, T)$ (removal times)
%			\ENDFOR
		\FOR {$k = 1, \dots, K$}
		    \STATE $\mu_k \leftarrow \beta I(t_{k-1})$
		    \FOR {$j \in \mathcal{I}_k$}
		        \STATE $z^I_j \sim \textsf{TrunExp}(\mu_k; t_{k-1}, t_k)$ (infection times)
		        \STATE $p_j \leftarrow P(z^R_j \le T | z^I_j)$ %1 - \exp\{-\lambda(T - z^I_j)^a\}$
		        \STATE $z^R_j|z^I_j \sim (1 - p_j) \delta_{\infty}(z^R_j) + p_j \textsf{TrunF}(z^R_j - z^I_j; \lambda, 0, T - z^I_j)$ (removal times)
		    \ENDFOR
%			    \STATE $R_k \leftarrow \#\{i:\tau^R_i \in (t_{k-1}, t_k]\}$ (number of removals in the $k^{\text{th}}$ interval)
%			    \STATE $I(t_k) \leftarrow I(t_{k-1}) + I_k - R_k$
		\ENDFOR
	\end{algorithmic}
\end{algorithm}

\subsection{Importance of the Surrogate Process}  \label{sec:imp}

The block sampler for the latent data can be said to be \textit{semi-independent}: by this we mean that the current and proposed latent data are independent conditionally on the current values of the parameters: $\mathbf{Z} \independent  \mathbf{Z}^\star | \theta$. This characteristic of the latent data sampler will be crucial for proving the uniform ergodicity of the resulting Markov chain (see Section \ref{sec:uni}) and contrasts with existing single-site samplers which update only a small fraction of the latent data per iteration and, as a result, generate consecutive configurations of the latent data that are mostly identical.
We therefore have a Gibbs-like sampler in which the parameters are updated conditionally on the latent data and the latent data are updated conditionally on the parameters only. If we were able to simulate directly form the SEM conditionally on the observed data, we would have an exact Gibbs sampler in which all proposed latent data are accepted. Since we instead simulate from a surrogate process, we need to accept the proposed latent data according to a M-H scheme to ensure that the Markov chain converges to the posterior distribution under the original SEM.

This semi-independence of the latent data sampler implies that the efficiency of the DA-MCMC algorithm to explore the latent space is directly related to the M-H acceptance rate of the proposed latent data, with a larger acceptance rate resulting in better mixing.
The acceptance rate in turn depends on how faithfully the proposal process resembles the target process. If the two processes are similar, then the proposed latent data will be accepted with high probability and the Markov chain will have excellent mixing properties.

\begin{figure}
	\centering
	\subfloat[$K = 5$   ]{ \includegraphics[width = .45\textwidth]{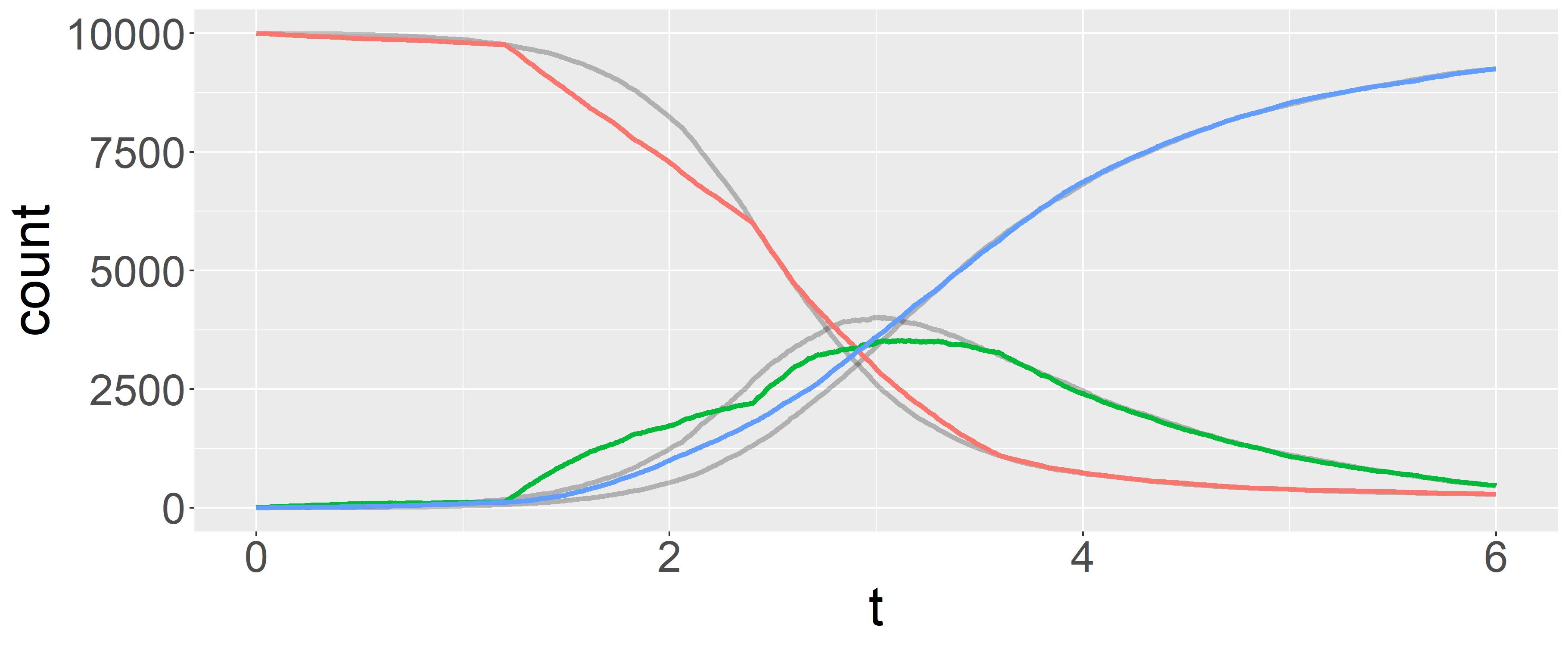}}
	\subfloat[$K = 10$  ]{ \includegraphics[width = .45\textwidth]{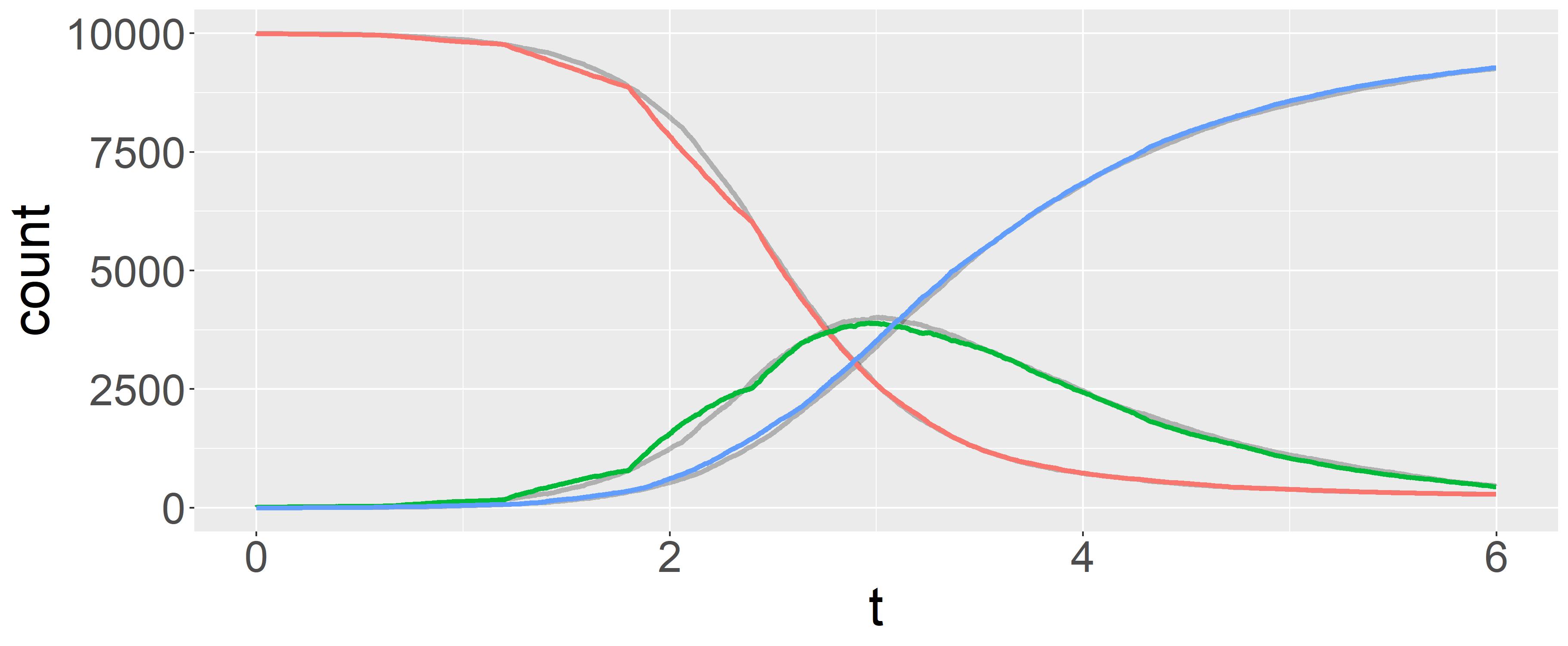}}\\
	\subfloat[$K = 50$  ]{ \includegraphics[width = .45\textwidth]{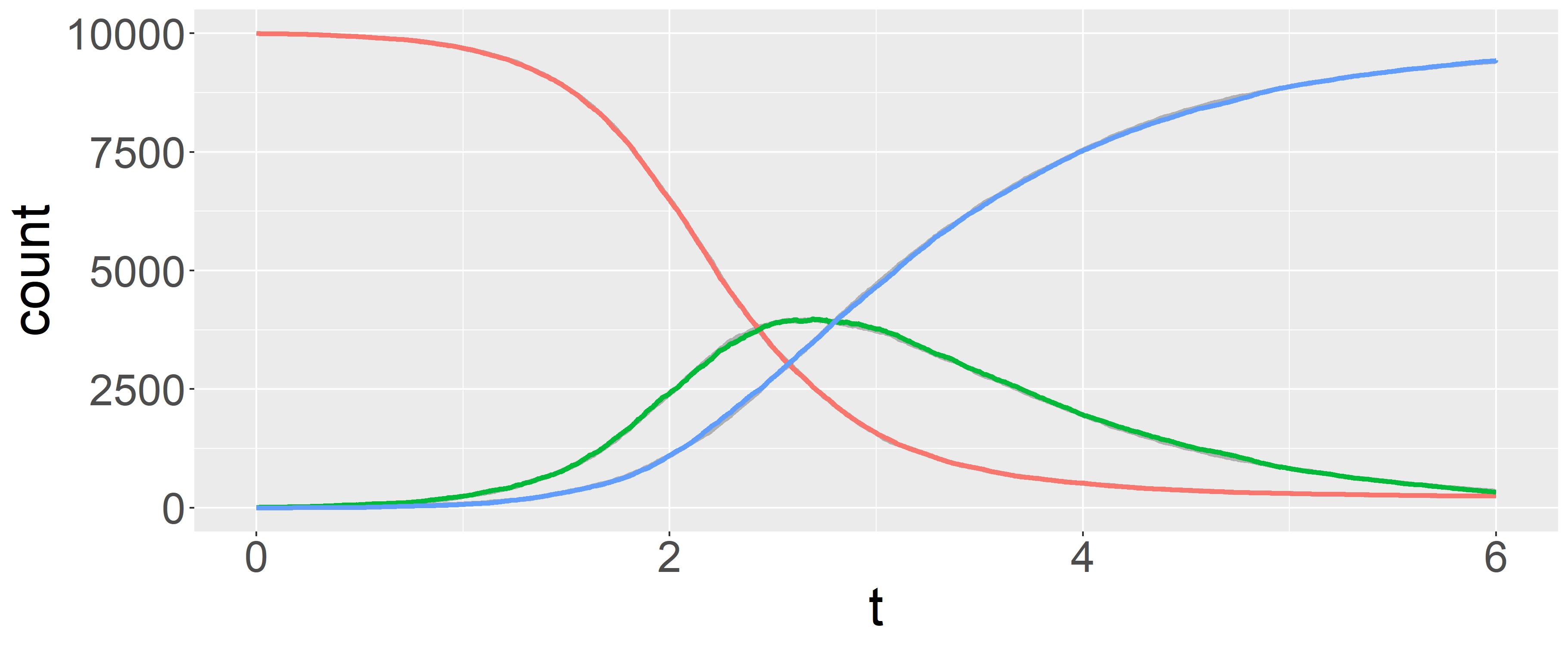}}
	\subfloat[$K = 1000$]{ \includegraphics[width = .45\textwidth]{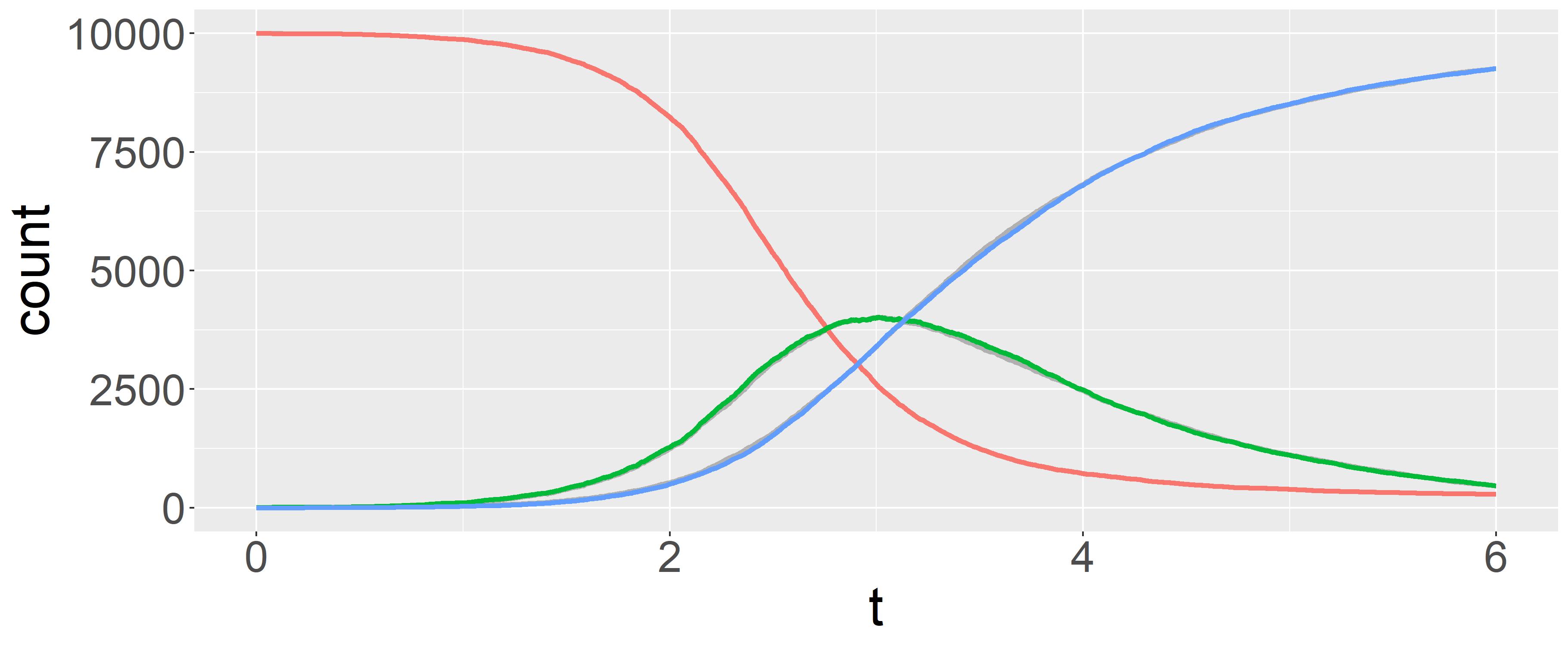}}
	
	\caption{Compartment trajectories in a SIR (in colors, with $S$ in red, $I$ in green and $R$ in blue) and a PD-SIR process (in grey) with the same incidence counts $(I_1, \dots, I_K)$.}
	\label{fig:comparison}
\end{figure}

In our case, the surrogate PD-SIR only differs from the SIR in its infection dynamics, with the removal dynamics being identical in the two processes. Figure \ref{fig:comparison} compares the trajectories of the compartments $S$, $I$ and $R$ of a SIR process of moderate size ($(S(0), I(0)) = (1000, 10)$ with $(\beta, \lambda, a) = (0.003, 1, 1)$ and $T = 6$) and those of four PD-SIR processes constrained to be consistent with the observed incidence data $(I_1, \dots, I_K)$ from the SIR process for $K \in \{5, 10, 50, 1000\}$. We see that the PD-SIR is qualitatively close to the SIR, even for small values of $K$. Unsurprisingly, the quality of the approximation improves as $K$ increases. In the limit as $K \rightarrow \infty$, the infection times are effectively known under the PD-SIR and the two processes become stochastically equivalent. The seemingly piece-wise linear trajectory of the $S$ compartment in the PD-SIR visible for small $K$ comes from it following a LDP with piece-wise constant death rate.
This observation that the PD-SIR faithfully approximates the SIR, even for moderate $K$, provides intuition for why our method can yield a high acceptance rate in the M-H step for the latent data. As a result, the Markov chain makes large jumps in the high-dimensional latent space and thus explores it efficiently.

For large outbreaks, however, the acceptance rate may drop considerably, thereby hindering the mixing of the Markov chain. To address this issue, we introduce a tuning parameter $\rho \in (0, 1]$ which determines the proportion of individuals whose trajectory is updated per iteration. Each iteration, the sampler updates the infection and removal times of only a subset of $\lceil\rho n_I\rceil$ individuals chosen uniformly at random.
A smaller $\rho$ will typically result in a larger M-H acceptance rate, which may lead to better overall mixing in large populations.

\subsection{Uniform Ergodicity}  \label{sec:uni}

We now turn to the analysis of the convergence properties of the Markov chain $\left\{\left(\theta^{(m)}, \mathbf{z}^{(m)}\right)\right\}_m$ underlying the MCMC algorithm.
%It is straightforward to show that the chain is ergodic (see Appendix \ref{app:erg}). By the ergodic theorem, the estimator $\bar{g}_m := \frac{1}{m} \sum_{i=0}^{m-1} g\left(\theta^{(m)}, \mathbf{z}^{(m)}\right)$ is therefore consistent for $E_\pi g$ whenever $E_\pi |g| < \infty$, for any starting value \citep{Tierney.1994}.
%Additionally, 
A Markov chain is said to be \textit{uniformly ergodic} if for some finite $M$ and positive constant $r<1$, the $n$-step transition kernel $P^n$ satisfies
$$
\Vert P^n(x,.)-\pi(.)\Vert_{TV} \le M r^n, \quad \forall x \in \chi
$$
where $\Vert \mu \Vert_{TV} := \sup_A \mu(A)$ denotes the total variation norm of a signed measure $\mu$.
Uniform ergodicity ensures the existence of a Central Limit Theorem for the estimator $\bar{g}_m := \frac{1}{m} \sum_{i=0}^{m-1} g\left(\theta^{(m)}, \mathbf{z}^{(m)}\right)$ of $E_\pi g$ whenever $E_\pi |g|^2 < \infty$ \citep{Jones.2004b}: for any initial distribution
$$
\sqrt{m}(\bar{g}_m - E_\pi g) \Rightarrow \textsf{N}(0, \sigma^2_g)
$$
as $m \rightarrow \infty$, where $\Rightarrow$ denotes convergence in distribution, and $\sigma^2_g$ is some positive, finite value which can be estimated via regeneration sampling, batch means or a spectral variance analysis \citep{Flegal.2010}.

\begin{theorem}  \label{theo:uni}
	%The transition kernel $P$ of the proposed DA-MCMC algorithm satisfies,
	%$$
	%P((\theta, \mathbf{z}), .) \ge \beta \nu(.) , \quad \forall (\theta, \mathbf{z}) \in \chi_\theta \times \chi_\mathbf{z}
	%$$
	%for some $\beta>0$ and probability measure $\nu$. As a result,
	The state space $\chi = \chi_{\theta}\times\chi_{\mathbf{z}}$ is a \textit{small set} for the transition kernel of the proposed DA-MCMC algorithm for the stochastic SIR with Weibull-distributed infection periods, which implies that the underlying Markov chain is uniformly ergodic.
\end{theorem}

Here $\chi_{\theta}$ denotes the parameter space of $\theta = (\beta, \lambda)$ and $\chi_{\theta}$ is defined in \eqref{eq:X}.
This result is noteworthy since requiring the entire space to be small is typically quite restrictive, and as a result often not satisfied in models with unbounded spaces.
The complete proof of Theorem \ref{theo:uni}, which takes advantage of the semi-independence of the latent data sampler, can be found in Appendix \ref{app:uni}. While the proof is tailored to Weibull-distributed infection periods, it easily generalizes to other distributions such as the gamma.

\section{Performance on simulated and real epidemic data}  \label{sec:per}
	
\subsection{Simulation Study}  \label{sec:sim}
%Outline: (1) proof of concept, (1bis) coverage, (2) rho (3) single-site versus joint proposal

We validate the mixing properties of the proposed DA-MCMC algorithm empirically via a suite of simulation studies of the non-Markovian stochastic SIR process with Weibull-distributed infection periods.
Throughout, we employ the conjugate priors \eqref{eq:pri_beta} and \eqref{eq:pri_lambda} (see Appendix \ref{app:wei}) with weakly informative hyper-parameters $\beta \sim \textsf{Ga}(0.01, 1)$ and $\lambda \sim \textsf{Ga}(0.01, 1)$ independently and fix the Weibull shape parameter to $a=2$.

%The basic reproduction number, which indicates the expected number of secondary infections caused by an infectious individual in a susceptible population, corresponds to $R_0=\beta*\iota$, where $\iota$ corresponds to the mean of the distribution $F$ and denotes the expected duration of the infection period.

\begin{figure}
	\centering
	\subfloat[Compartments ($S$ in red, $I$ in green and $R$ in blue)]{\includegraphics[width = .32\textwidth]{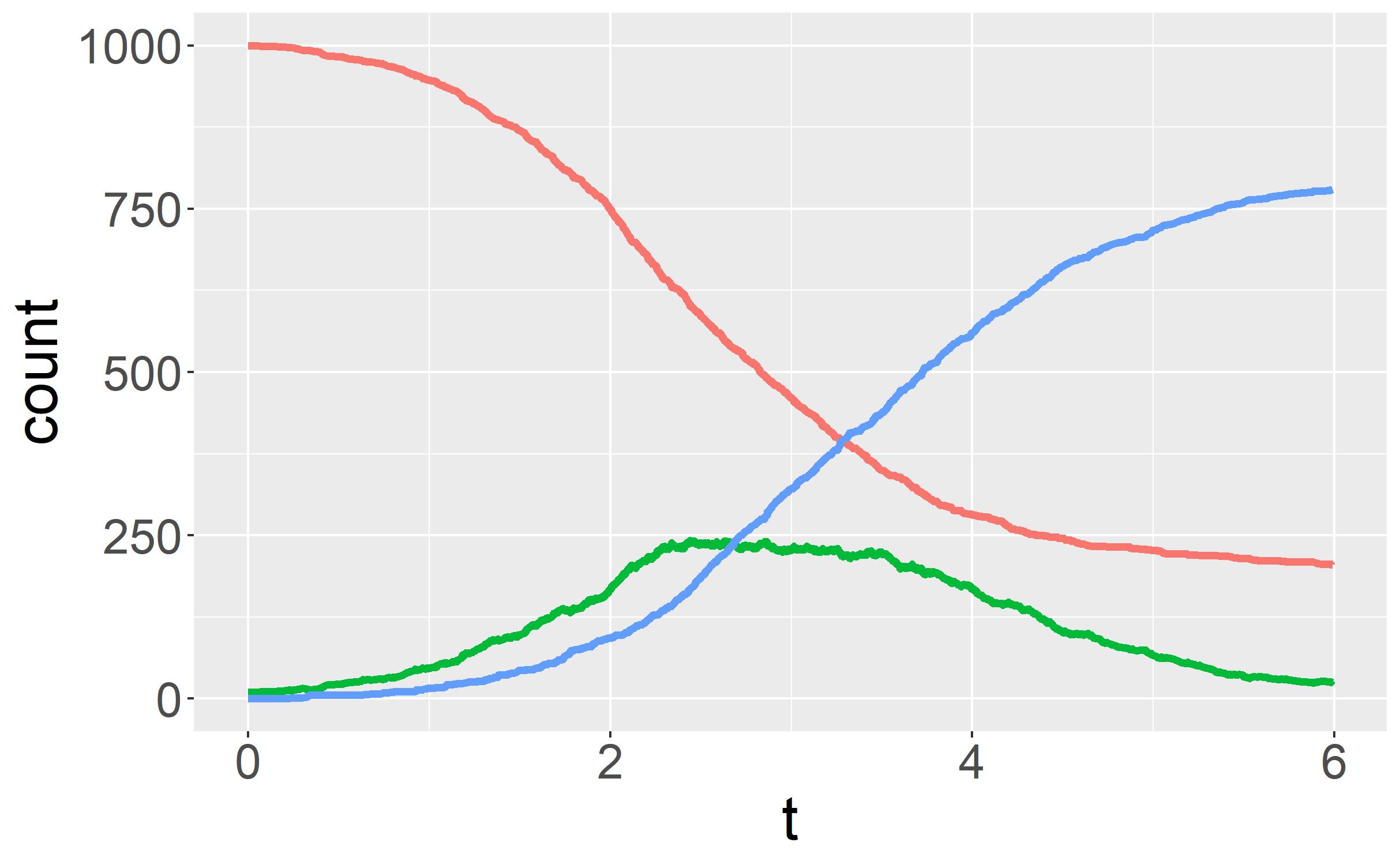} \label{fig:E1_trajectories}}
	\subfloat[Traceplot during the transient phase                   ]{\includegraphics[width = .32\textwidth]{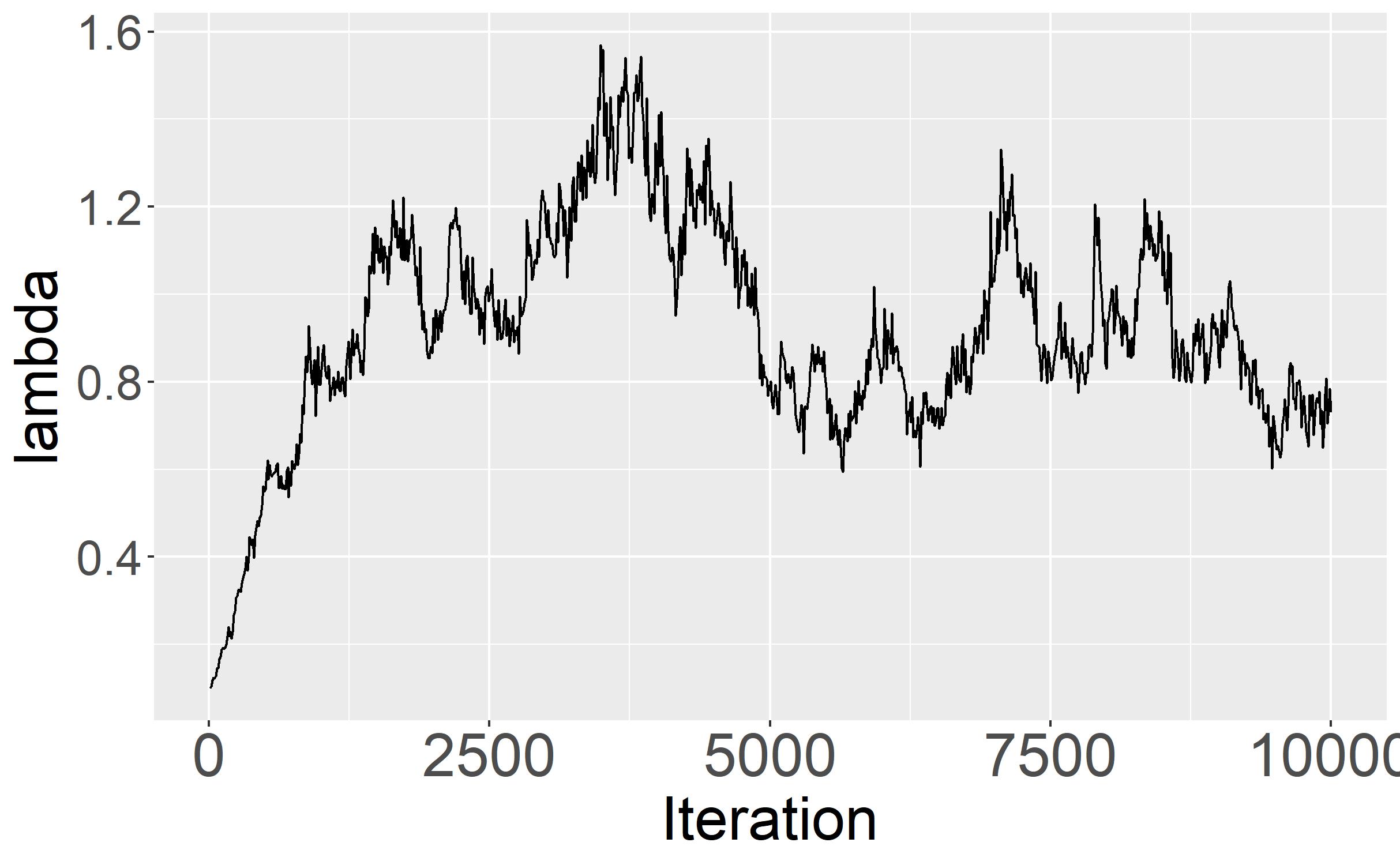} \label{fig:E1_short_no_burn_lambda_tp}}
	\subfloat[Traceplot after the transient phase                    ]{\includegraphics[width = .32\textwidth]{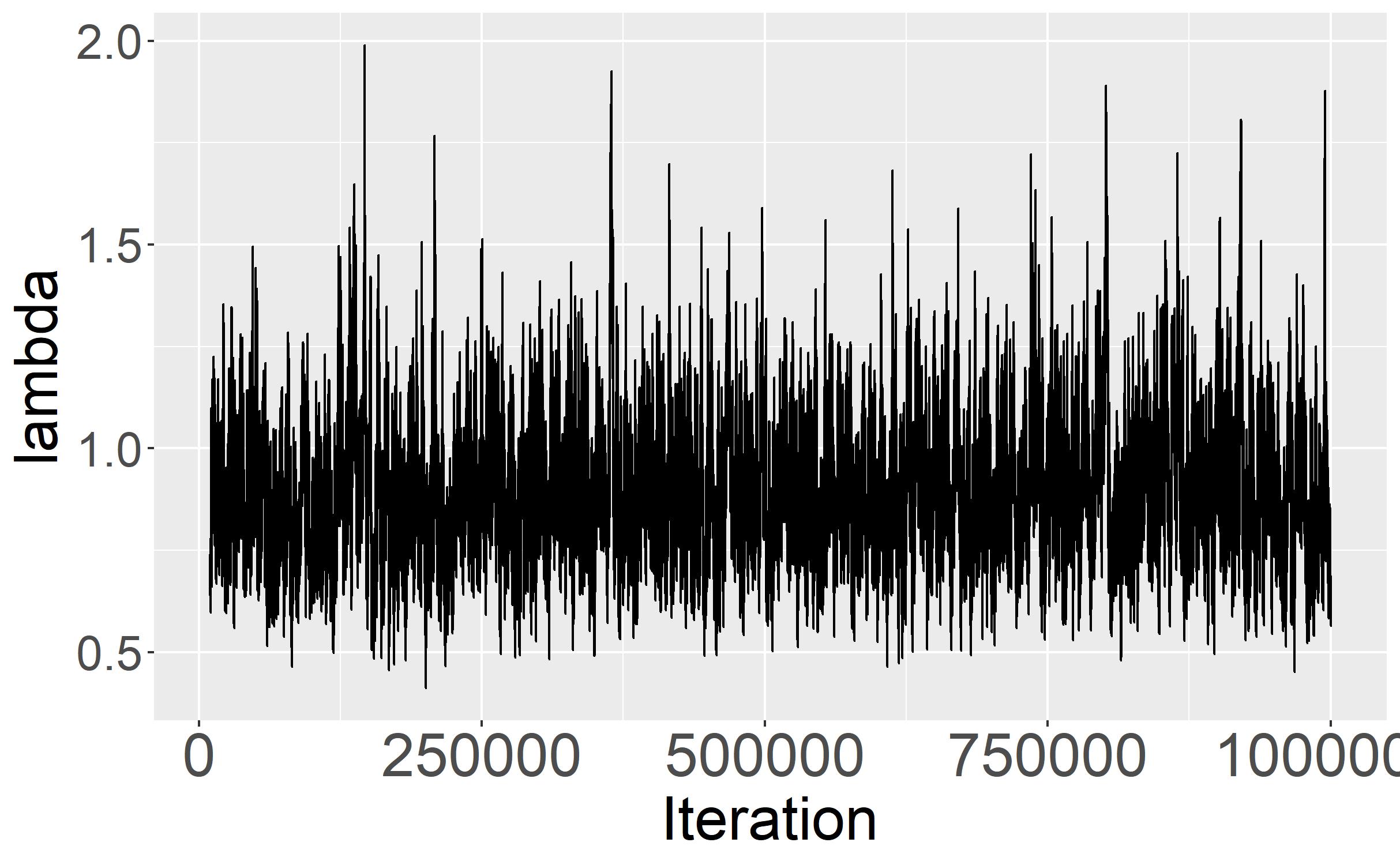}\label{fig:E1_burn_lambda_tp}}
	
	\caption{Performance of DA-MCMC in a medium-sized population for the parameter $\lambda$.}
	\label{fig:E1}
\end{figure}

First, we assess the convergence and mixing of the algorithm in a moderately sized population of 1,000 individuals. We simulate an epidemic with true parameters $(\beta, \lambda) = (0.00225, 1)$ --- giving the basic reproduction number $R_0 = \beta \lambda^{1/a} / \Gamma(1+1/a) = 2$ --- starting with $(S(0), I(0)) = (1000, 10)$ until time $T = 6$, when the outbreak has completed most of its course but is not over yet $(I(T) = 25)$, as depicted in Figure \ref{fig:E1_trajectories}. The numbers of infections in $K = 10$ equal-length time intervals are observed, corresponding to $\mathbf{Y} = (
12, 13, 21, 46, 91, 127, 156, 151, 88, 41)$, which gives a total of $n_I = 746$ infections. To our knowledge, this is the largest epidemic for which exact Bayesian inference has been conducted on a SEM.

To evaluate the convergence of the Markov chain, we initialize it in a low density region at $\left(\beta^{(0)}, \lambda^{(0)}\right) = (0.000225, 0.1) = \left(\beta/10, \lambda/10\right)$. Note that a practical benefit of the semi-independent sampler is that we can initialize the Markov chain by only specifying initial values $\theta^{(0)}$ for the parameters since $\mathbf{Z}^{(0)}$ can be generated from $\theta^{(0)}$ alone.
We keep every $10$th draw due to storage and set $\rho = 0.2$, updating the trajectories of a random subset of $\lceil\rho n_I\rceil = 159$ individuals each iteration. This yields a healthy acceptance rate of $0.21$.

Remarkably, the algorithm takes less than $30$ minutes to run $1$ million iterations on a personal laptop. 
Figure \ref{fig:E1_short_no_burn_lambda_tp} shows the traceplot of the parameter $\lambda$ during the transient phase. We see that the Markov chain rapidly migrates from the low density region where it started to the mode of the target distribution around the value $1$ in 1,000 iterations. Figure \ref{fig:E1_burn_lambda_tp} shows that once the chain reaches the high density region it mixes well.
The posterior means of $\beta$, $\lambda$ and $R_0$ are respectively $0.00214$, $0.894$ and $2.02$, respectively based on $813$, $698$ and 4,806 effective sample sizes. The $90\%$ equal-tailed Bayesian credible intervals (BCI) are $(0.00186, 0.00245)$, $(0.642, 1.20)$ and $(1.88, 2.17)$ and cover the true values of the parameters.
	
Second, we validate the uncertainty quantification of the posteriors under our algorithm over repeated simulations. To this end, we examine the frequentist coverage properties of the $90\%$ equal-tailed BCI. We repeat the previous simulation 2,000 times and compute the posterior mean of the parameters and the corresponding credible intervals. Since the priors are only weakly informative, the observed coverage rate of the BCI should be close to $90\%$.
Table \ref{tab:coverage} provides the empirical coverage rate of the BCIs for the parameters $\beta$, $\lambda$ and $R_0$ along with the average and variance of the posterior means. As expected, the BCIs cover the true values around $90\%$ of the time, well within Monte Carlo error of the nominal coverage rate under a weakly informative prior. This result suggests that running the Markov chain for $1$ million iterations suffices to approximate the posterior distribution of the parameters.

\begin{table}
	\caption{Empirical coverage of $90\%$ BCI and summary statistics of posterior means across 2,000 independent runs in a medium-sized population ($n=1000$). The true values of the parameters are $(\beta, \lambda, R_0) = (0.0025, 1, 2.82)$. %The standard errors for the observed coverage rate are calculated as $\sqrt{\hat{p}(1-\hat{p})/2000}$, which results in a largest standard error of $6.8$e-3.}
	\label{tab:coverage}}
	\centering
	\fbox{%
	\begin{tabular}{ C{2cm}| *{3}{C{3cm}}}
		%\hline
		Parameter & Observed coverage rate & Average of the posterior means & SD of the posterior means \\ 
		\hline
		$\beta$ & 0.895 & 0.0025 & 0.000255 \\ 
		$\lambda$ & 0.902 & 1.00 & 0.142 \\ 
		$R_0$ & 0.910 & 2.84 & 0.194 \\
		\hline
	\end{tabular}
	}
\end{table}

\begin{figure}
	\centering
	\subfloat[Run time in minutes]{
	    \includegraphics[width = .3\textwidth]{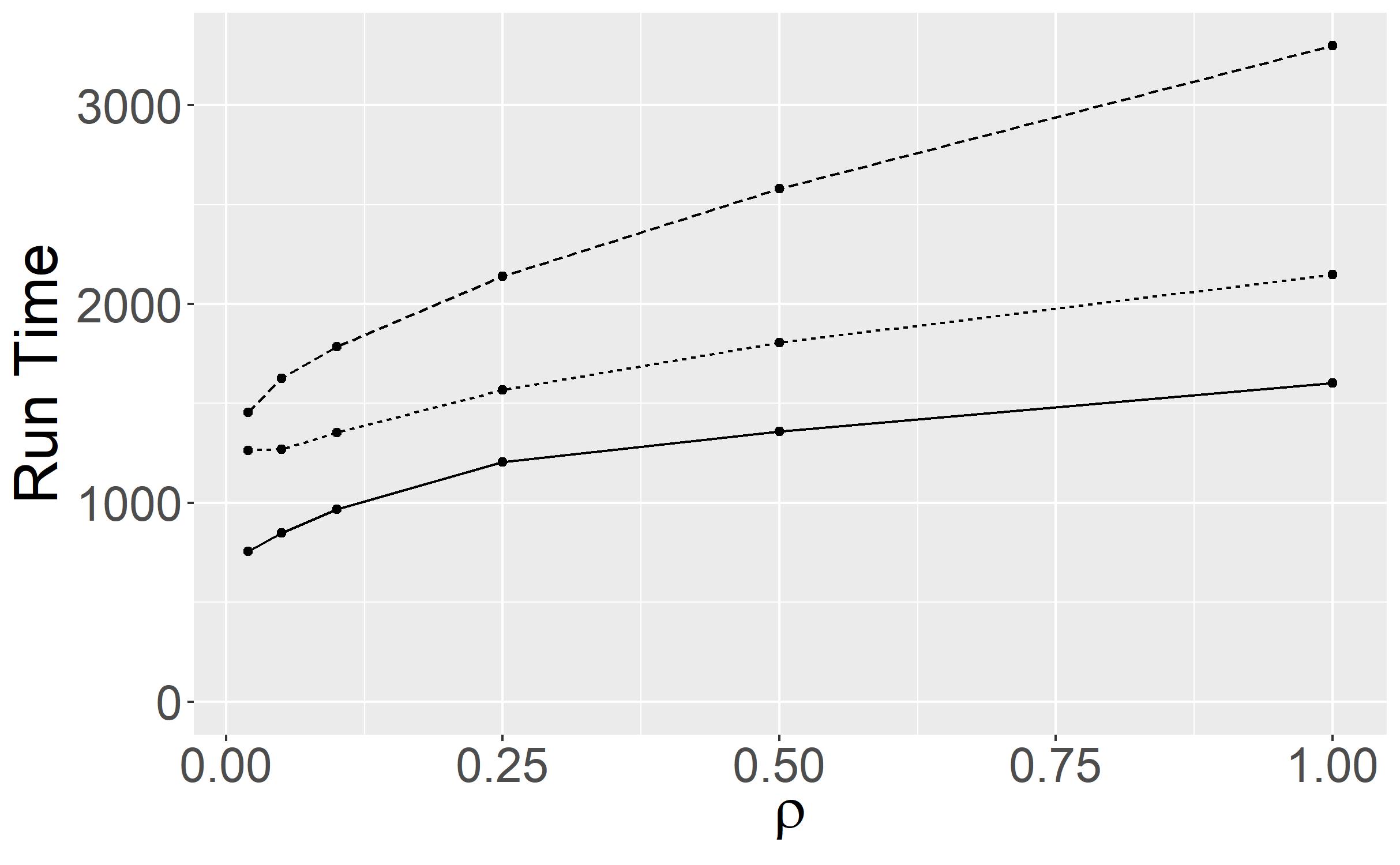}
	\label{fig:E3_runtime}}
	\subfloat[Acceptance rate    ]{
	    \includegraphics[width = .3\textwidth]{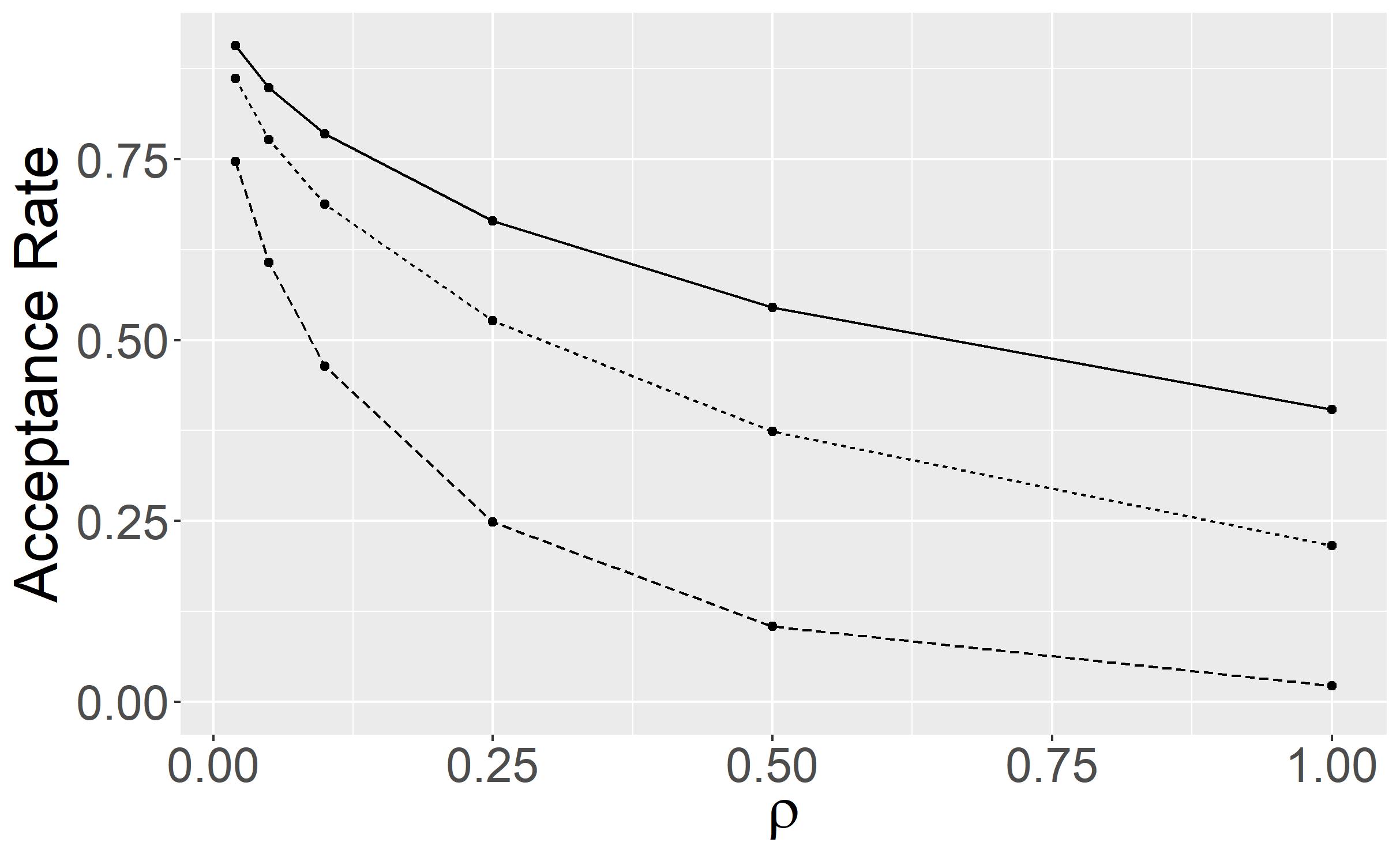}
	    \label{fig:E3_accept}}
	\subfloat[ESS/sec for $R_0$  ]{
	    \includegraphics[width = .3\textwidth]{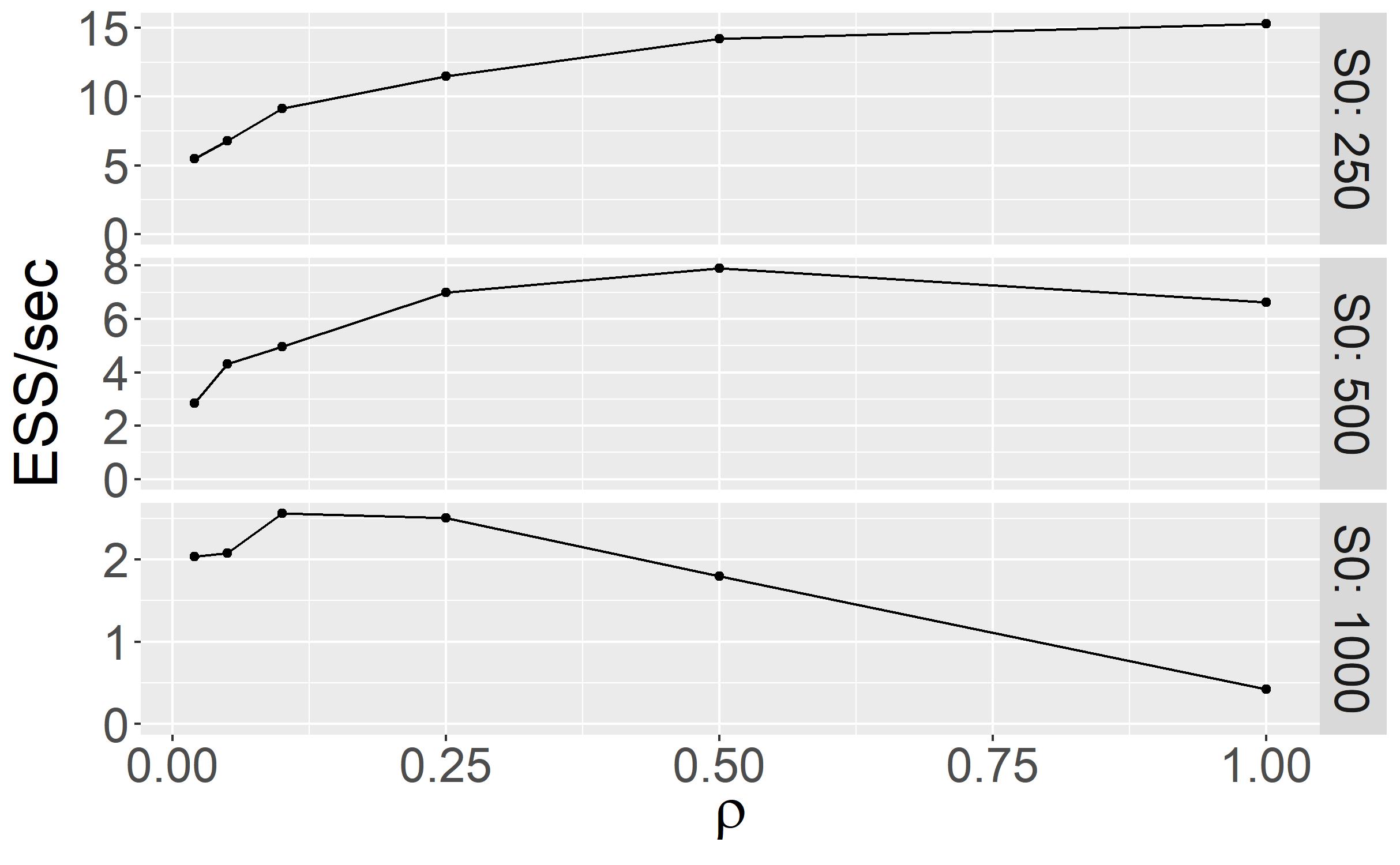}
	    \label{fig:E3_facet_ESSsecR0}}
	    
	\caption{Impact of $\rho$ on the performance of the DA-MCMC in a population of size $250$ (solid), $500$ (dotted) and $1000$ (dashed).}
	\label{fig:E3}
\end{figure}

Third, we explore the impact of the tuning parameter $\rho$ on the mixing of the MCMC as the population size varies.
We simulate three realizations from the SIR process with three population sizes $S(0) \in \{250, 500, 1000\}$, and $I(0)=10$, $\lambda=1$, $T = 6$ and $K = 10$ for each simulation. We choose $\beta$ so that $R_0 \in \{2.2, 2.5, 3\}$ respectively, in order to obtain three outbreaks that have completed most of their course but are not over yet. For each population size and each value $\rho \in \{0.02, 0.05, 0.1, 0.25, 0.5, 1\}$, the MCMC algorithm is run for $1$ million iterations. We use the effective sample size per second (ESS/sec) to compare the mixing of the resulting Markov chains.
Figures \ref{fig:E3_runtime} and \ref{fig:E3_accept} respectively show the running time in seconds of the algorithm and the acceptance rate in the M-H step. As expected, the run time increases with the population size $n$ and with the tuning parameter $\rho$, which reflects the fact that the number of random variables to generate per iteration is proportional to $\lceil\rho n_I\rceil$. Moreover, for each population size, the acceptance rate decreases as the number of trajectories updated per iteration increases.
Finally, the effect of $\rho$ on the mixing properties of the Markov chain is presented in Figure \ref{fig:E3_facet_ESSsecR0}. For a population of $250$ individuals, updating the entire latent data in the M-H step provides the best mixing. As the population size increases, however, the optimal value of $\rho$ decreases. For instance, for a population of 1,000 individuals, updating the whole latent data results in an excessively low acceptance rate, while updating less than $10$\% of the latent data makes the updates too small to efficiently explore the latent space.

\begin{figure}
	\centering
	\subfloat[block sampler]{
		\includegraphics[width = .24\textwidth]{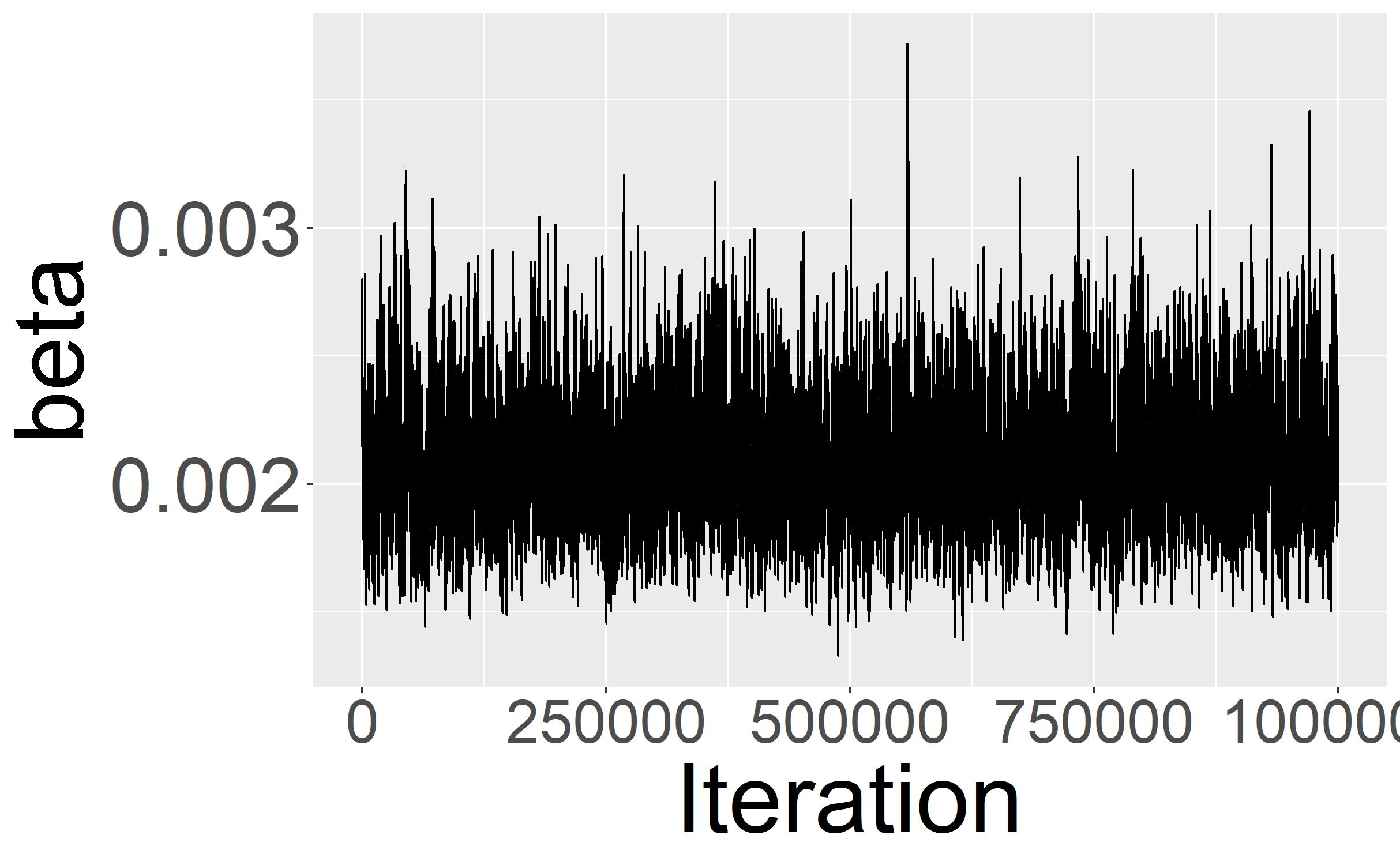}
		\includegraphics[width = .24\textwidth]{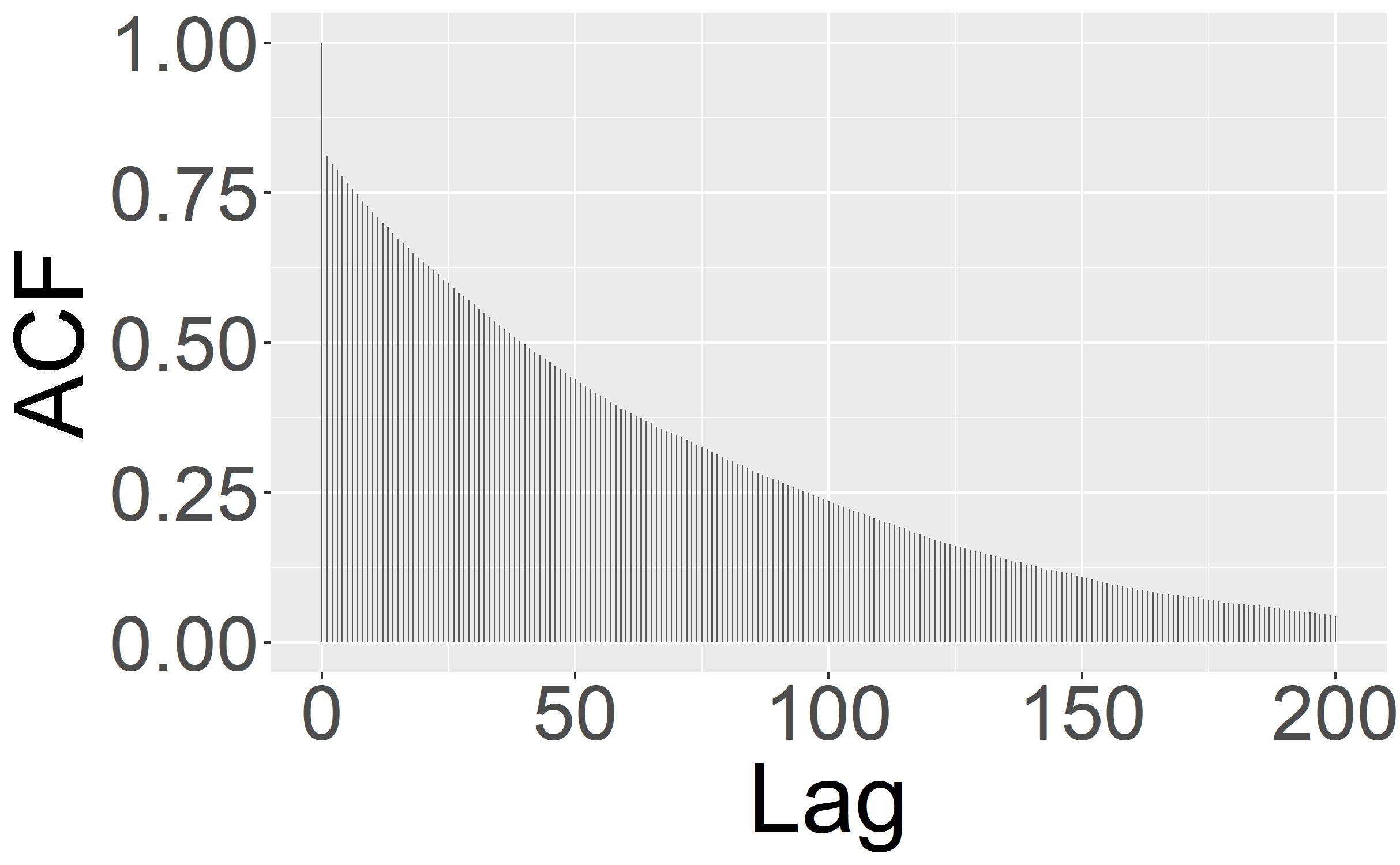}
		\label{fig:E6_no_burn_beta_tp_joint}}
	\subfloat[single-site sampler]{
		\includegraphics[width = .24\textwidth]{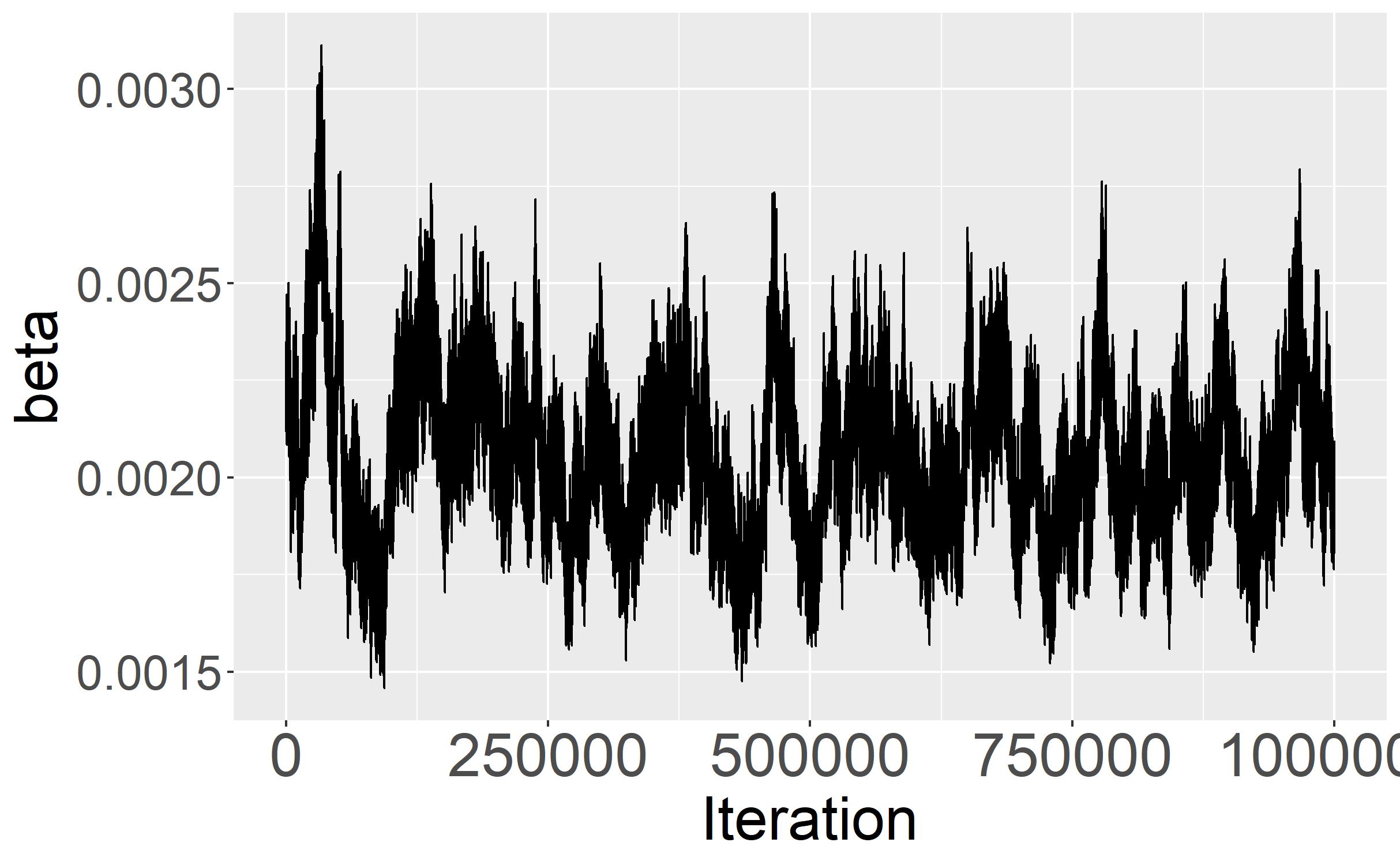}
		\includegraphics[width = .24\textwidth]{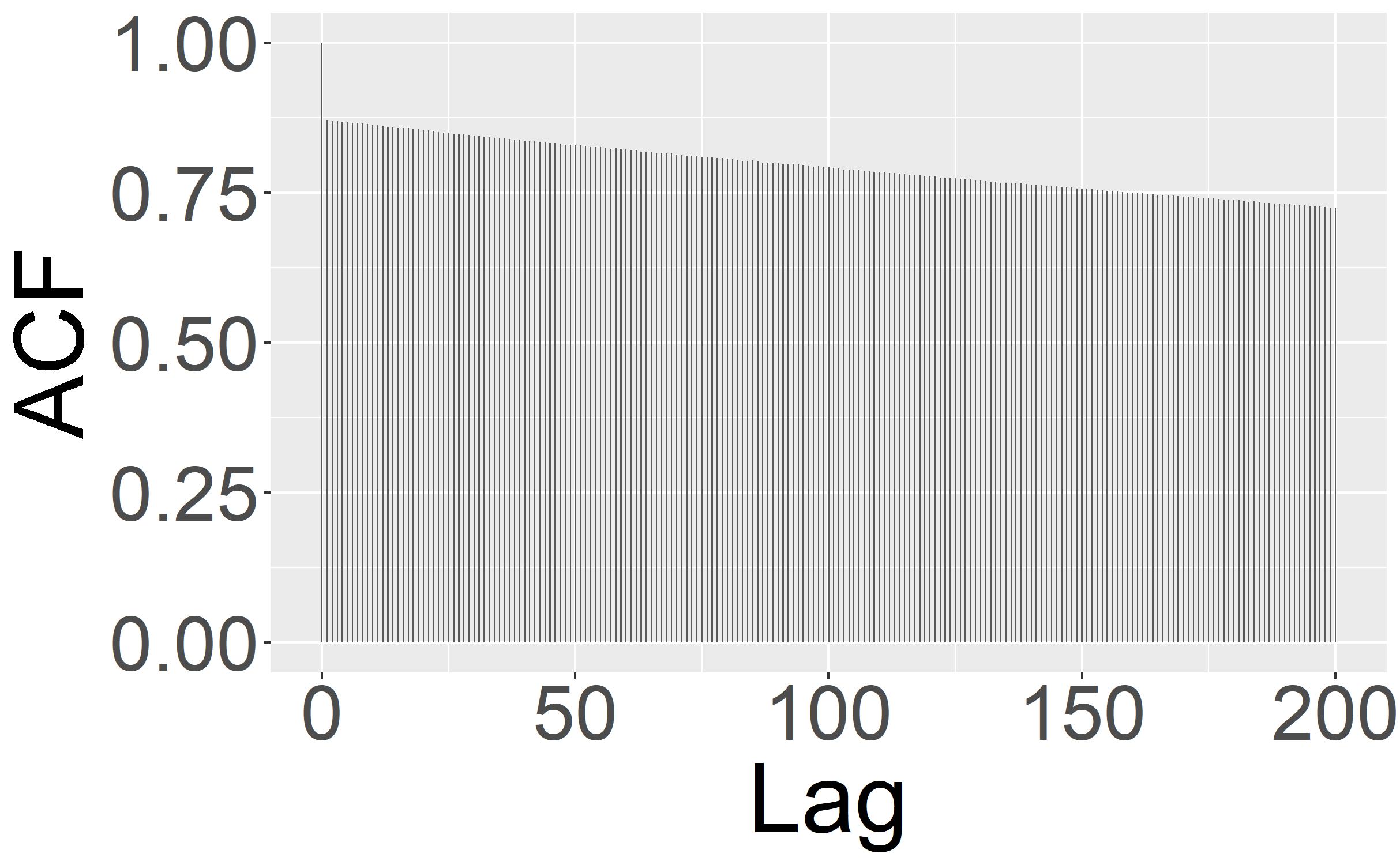}
		\label{fig:E6_no_burn_beta_tp_single}}
	
	\caption{Traceplots and auto-correlation functions of the block and single-site samplers for $\beta$.}
	\label{fig:E6}
\end{figure}

Finally, we showcase the efficacy of our block sampler for the latent data through a comparison %the performance of the proposed DA-MCMC algorithm in which event times are jointly proposed 
to a single-site sampler that is similar in spirit to those in \cite{Gibson.1998}, \cite{ONeill.1999} and \cite{Fintzi.2017}. We do not consider the particle filters of \cite{King.2015} nor the diffusion approximations of \cite{Fintzi.2020} since these methods are not suited to situations in which incidence counts are observed without noise; see \cite{Ho.2018}.
The single-site sampler that we consider here is a special case of our block sampler in which $\rho = n^{-1}$, so that the infection and removal times of a single individual are updated each iteration. The algorithms are run on simulated data for a population of 1,000 individuals for $1$ million iterations. In the block sampler, $\rho$ is set to $0.1$. Figure \ref{fig:E6} presents the traceplots for $\beta$ as well as the auto-correlation functions for the two algorithms. The Markov chain mixes much better when event times are jointly proposed. This is supported quantitatively in Table \ref{tab:E6} which shows that the ESS/sec obtained with the block sampler is between 10 and 16 times larger than that of the single-site sampler. Hence, even though our block sampler is computationally more costly per iteration, it is more efficient overall.

\begin{table}
	\caption{ESS/sec for the block and single-site samplers.}
	\label{tab:E6}
	\centering
	\fbox{\begin{tabular}{ C{2cm}| *{2}{C{3cm}}}
		Parameter & block sampler & single-site sampler \\ 
		\hline
		$\beta$ & 0.48 & 0.055 \\ 
		$\lambda$ & 0.44 & 0.045 \\ 
		$R_0$ & 3.4 & 0.21 \\
		\hline
	\end{tabular}
}
\end{table}

\subsection{Ebola Outbreak in Gu\'eck\'edou, Guinea}  \label{sec:ebo}
%Outline: background; setup; results; table; figure

We now turn to a case study concerning the 2013-2015 Ebola outbreak in Western Africa.
Between late $2013$ and $2015$, Guinea, along with several neighboring countries, experienced the largest outbreak of the Ebola virus disease in history. The virus, which has a fatality rate of $70\%$, was responsible for the death of almost $2000$ people in Guinea alone.
\cite{Coltart.2017} traced back the origin of the outbreak to the Gu\'eck\'edou prefecture, Guinea, at the end of November $2013$. Weekly counts of positive tests, which we model as infection counts in the SIR, are available for each prefecture for the $73$ weeks between the end of December $2013$ and May $2015$.

\begin{figure}
	\centering
	\subfloat[Weekly infection counts]{
	    \includegraphics[width = .32\textwidth]{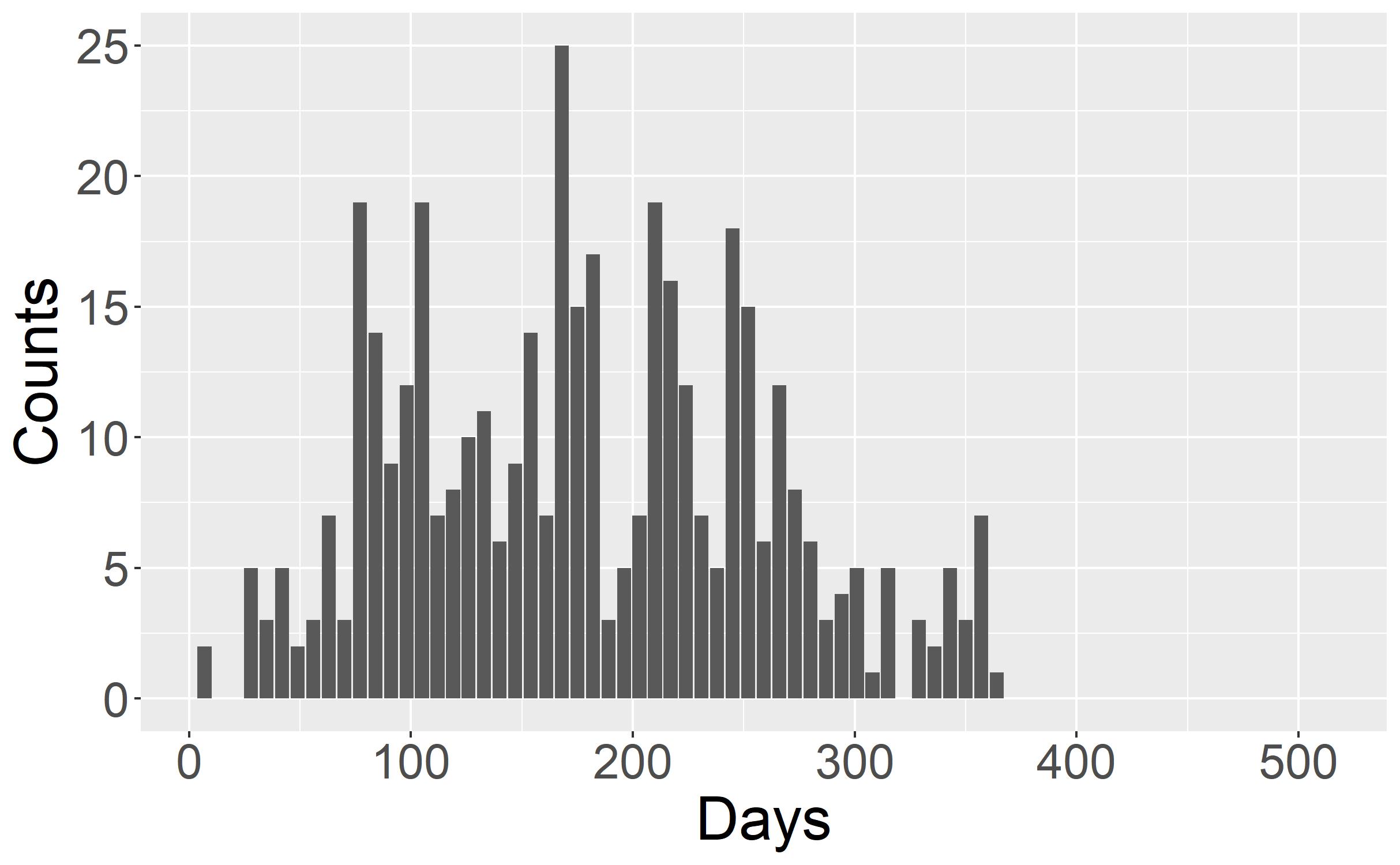} 
	    \label{fig:E5_observed_data}}
	\subfloat[Posterior distribution]{
	    \includegraphics[width = .32\textwidth]{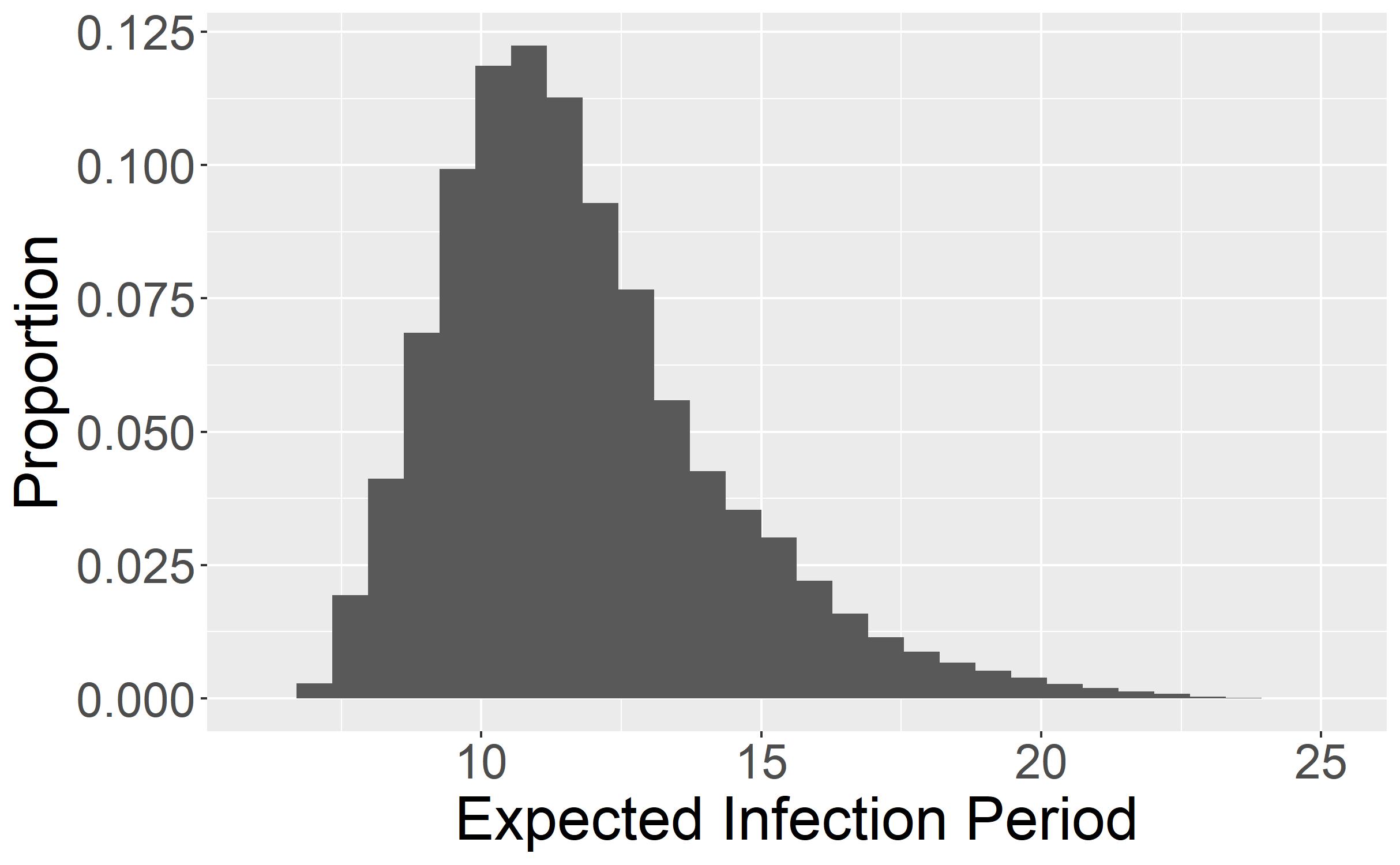} 
	    \label{fig:E5_lambda_hist}}
	\subfloat[Traceplot]{
	    \includegraphics[width = .32\textwidth]{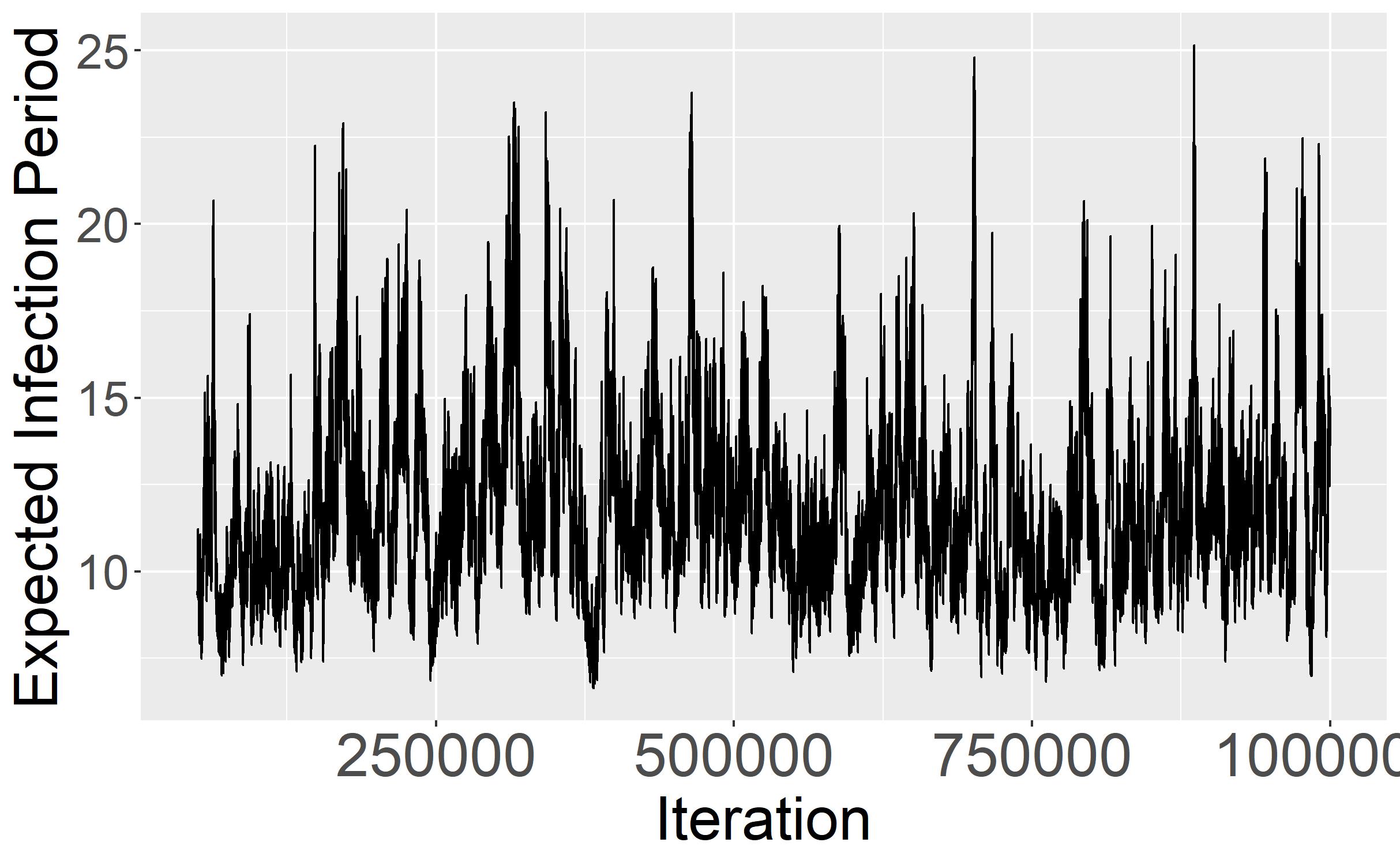}
	    \label{fig:density_R0}}
	    
	\caption{Weekly infection counts observed in Gu\'eck\'edou, and the posterior distribution and traceplot of the expected infection period.}
	\label{fig:ebola}
\end{figure}

We fit the stochastic SIR model with Weibull-distributed infection periods to the incidence counts of the Gu\'eck\'edou prefecture with our block DA-MCMC algorithm.
It must be pointed that the SIR model is only illustrative and serves to show that fast and exact Bayesian inference can be made in a large population with the proposed algorithm rather than providing new insights into this outbreak or the Ebola virus in general.
The units of time are ``days", and $t=0$ corresponds to Monday December 30, 2013, the first week for which infection counts are available. The timeline of the $410$ infections that were observed in Gu\'eck\'edou is shown in Figure \ref{fig:E5_observed_data}.
As the first documented infection occurred in late November 2013 \cite{Coltart.2017}, one month before the first reported infection count, we set $I(0) = 5$.
We further set the population size to $n = 292000$, the estimated number of people living in the Gu\'eck\'edou prefecture in $2014$, and employ the weakly informative conjugate priors of Section \ref{sec:sim}. Finally, we set the Weibull shape parameter to $a=2$.

We initialize the Markov chain at $(\beta^{(0)}, \lambda^{(0)}) = (10^{-7}, 0.05)$ and run it for $1$ million iterations, updating the trajectories of $\rho=10\%$ of the individuals each iteration. %Even for such a large population, the total run time of the algorithm was less than $50$ minutes on a personal laptop.
The M-H step for the latent data proposals achieves a healthy $20.1\%$ acceptance rate. The first 50,000 iterations of the Markov chain are discarded as a burn-in.
Figure \ref{fig:ebola} shows the marginal posterior distribution of the expected infection period  $\lambda^{-1/a}\Gamma(1+1/a)$. The results suggests that individuals remained infectious for around $9$ days on average, which is consistent with the existing literature \cite{Coltart.2017}.

\section{Discussion and Conclusion}  \label{sec:dis}

The proposed block sampler makes the classical Metropolis-Hastings algorithm an efficient approach to fit stochastic epidemic models to incidence data. In contrast to methods that rely on approximate computation \citep{McKinley.2018} or on simplifying assumptions necessitated by computational considerations \citep{Fintzi.2020}, our data-augmented algorithm enables exact and fast Bayesian inference, even for large outbreaks, and leverages a well-studied, transparent MCMC framework with guarantees of uniform ergodicity.

Central to the success of the DA-MCMC algorithm is the efficient block sampler that we have designed for the latent data, which swiftly explores the latent space of epidemic paths consistent with the observed data. The PD-SIR process which we use to jointly propose latent variables possesses three features that make the sampler efficient. First, the PD-SIR closely approximates the SIR: the removal dynamics are identical and, for short observation time intervals, the infection dynamics are also very similar in the two processes. This enables the semi-independent block sampler to update a large portion of the latent data each iteration while maintaining a healthy acceptance rate. As a result, the Markov chain frequently makes large jumps in the latent space and has very good mixing properties. In contrast, existing DA-MCMC algorithms \citep{Gibson.1998, ONeill.1999, Fintzi.2017} keep most of the latent space fixed across iterations, which results in Markov chains that mix much more slowly.

Second, generating a PD-SIR process is extremely fast; it only requires simulations from the truncated exponential and $\mathcal{F}$ distributions, which can efficiently be realized via the inverse CDF method in the case of Weibull-distributed infectious periods. Moreover, ensuring that the generated latent variables are consistent with the observed data is done at no additional cost; in contrast, directly generating a SIR process consistent with the observed infection counts would be prohibitively slow \citep{Hobolth.2009}. 
Together, these two features of the PD-SIR make the DA-MCMC algorithm extremely fast and scalable to populations with hundreds of thousands of individuals.
%Moreover, the piece-wise constant infection rate of the PD-SIR is only updated $K$ times, as opposed to after each event in the SIR model,

%While existing data-augmented MCMC approaches have only be applied to populations of a few hundred of individuals, the analysis of the Ebola outbreak in Gu\'eck\'edou shows that our algorithm can be applied to populations of up to 300,000 individuals in a reasonable amount of time.	

Our algorithm features a tuning parameter $\rho$ that determines the proportion of individuals whose trajectory is updated per iteration, which in turn affects the acceptance ratio in the M-H step. Larger values for $\rho$ result in larger steps in the latent space but may give an excessively low acceptance rate, while smaller values of $\rho$ result in a higher acceptance rate but constrain the chain to make small jumps. Depending on the size of the population, different values of $\rho$ are optimal, and future work may seek theoretical insights to this end. In practice, one can use several short runs of the algorithm with different values for $\rho$ and select the value that yields the largest effective sample size per second.

%Exploring the latent space has been the computational bottleneck of DA-MCMC for stochastic epidemic models. The proposed method allows for efficient exploration of this space.

The DA-MCMC algorithm proposed in this article is specific to the stochastic non-Markovian SIR process, an arguably simplistic model for an outbreak of the Ebola virus.
Our data-augmentation framework and the idea of a faithful surrogate proposal can be extended to models of increasing realism. It will be fruitful to consider extensions to epidemic models accounting for a latency period (e.g., SEIR), under-reporting \citep{Fintzi.2017,Morozova.2021}, non-homogeneous mixing \citep{Severo.1969,Lomeli.2021} as well as a time-varying infection rate \citep{Kypraios.2018} in future work. We also invite readers to consider applications of these ideas in other stochastic process models with complex latent spaces that may similarly be made navigable through a well-designed surrogate proposal process embedded in a M-H algorithm.

%\bigskip
\begin{center}
{\large\bf SUPPLEMENTARY MATERIAL}
\end{center}

\begin{description}

\item[R Markdown] file to run the experiments reported in Sections \ref{sec:sim} and \ref{sec:ebo} (zip file)

\item[R package] PDSIR (https://github.com/rmorsomme/PDSIR) contains the code necessary to run the block DA-MCMC algorithm introduced in this article.

\item[Ebola data set] used in Section \ref{sec:ebo} (zip file)

\end{description}

\bibliographystyle{rss}
\bibliography{main}

\newpage
\appendix
\begin{center} \Large{\bf Appendix}	\end{center}

\section{Inference with Weibull-distributed infectious periods}  \label{app:wei}

If we let the infectious periods $\{\tau^R_i - \tau^I_i\}_{i\in \mathcal{I}}$ follow independent $\textsf{Wei}(\lambda, a)$ distributions with scale parameter $\lambda$, shape parameter $a$ and CDF $F(x)=1-\exp\left\lbrace-\lambda x^a\right\rbrace$, then the complete data likelihood \eqref{eq:cdl} becomes
\begin{align}
	L(\theta; \mathbf{X})
	& = \beta^{n_I} \lambda^{n_R} a^{n_R} \prod_{j \in \mathcal{I}} I(\tau^I_j) \exp\left\lbrace - \beta \int_{0}^{T} S(t)I(t)dt \right\rbrace 
	\prod_{k \in \mathcal{R}} (\tau^R_k - \tau^I_k)^{a-1} \nonumber \\
	\label{eq:cdl_wei}
	& \qquad \times \exp\left\lbrace - \lambda \left[ \sum_{k \in \mathcal{R}} (\tau^R_k - \tau^I_k)^a + \sum_{l \in \mathcal{R}^c} (T - \tau^I_l)^a\right] \right\rbrace
\end{align}
which belongs to the exponential family.

The gamma distribution is a conjugate distribution for the parameter $\lambda$ for the likelihood $\eqref{eq:cdl_wei}$. If we let
\begin{equation*}
	\lambda \sim \textsf{Ga}(a_{\lambda}, b_{\lambda}),
\end{equation*}
independently of $\beta$, then the full conditional distribution of $\lambda$ is
\begin{equation*}
	\lambda | \mathbf{X}, \beta \sim \textsf{Ga}\left( a_{\lambda} + n_R, b_{\lambda} + \sum_{k \in \mathcal{R}} (\tau^R_k - \tau^I_k)^a + \sum_{l \in \mathcal{R}^c} (T - \tau^I_l)^a \right)
\end{equation*}
Note that this distribution does not dependent on $\beta$.

Finally, it is straightforward to sample from the truncated Weibull distribution via the inverse CDF method. Let $U\sim U(0,1)$ be a uniform random variable between $0$ and $1$, then
\begin{equation*}
    X = (-\log(A-B*U)/\lambda)^{1/a}
\end{equation*}
with $A = \exp\{-\lambda l^a \}$ and $B = \exp\{-\lambda l^a \} - \exp\{-\lambda u^a \}$ follows a Weibull distribution with scale $\lambda$, shape $a$, bounded between $l$ and $u$. Generating truncated exponential random variables corresponds to the special case $a=1$.

\section{Proof of Theorem \ref{theo:ldp}}  \label{app:ldp}

Theorem \ref{theo:ldp} was first proved by \cite{Neuts.1971}. \cite{Ross.1996} provide a simpler proof which we now give.

\begin{proof}
	Consider a linear pure death process with $n$ particles and individual death rate $\mu$.
	Let $T_i$ be the time of the $i$th death. Then $W_1 = T_1 \sim \textsf{Exp}(n\mu)$ and $W_i = T_i - T_{i-1} \sim \textsf{Exp}((n-i)\mu)$ independently. Let $N$ be the number of deaths by time $t$. Then, 
	\begin{align*}
		& f(T_1 = t_1, \dots, T_N = t_N | N) \\
		& \propto f(T_1 = t_1, \dots, T_N = t_N, T_{N+1} > t) 1\{t_N < t\}\\
		& \propto f(W_1 = t_1, W_2 = t_2 - t_1, \dots, W_N = t_N - t_{N-1}, W_{N+1} > t - t_N) 1\{t_N < t\}\\
		& \propto \exp\{-n\mu t_1\} \exp\{-(n-1)\mu(t_2-t_1)\}\dots \exp\{-(n-N)\mu(t - t_N)\} 1\{t_N < t\}\\
		& \propto \exp\{-\mu t_1\}\exp\{-\mu t_2\} \dots \exp\{-\mu t_N\} 1\{t_N < t\}
	\end{align*}
	which corresponds to the kernels of independent exponential distribution truncated above by $t$. By the memoryless property of the exponential distribution, this results can be extended from the interval $(0, t]$ to any interval $(a, b]$.
\end{proof}

\section{Proof of Ergodicity}  \label{app:erg}

\begin{theorem}  \label{theo:erg}
	The Markov chain underlying the DA-MCMC described in this article algorithm is ergodic and Harris recurrent.
\end{theorem}

\begin{proof}
By construction, the distribution $\pi(\theta, \mathbf{z}|\mathbf{Y})$ is invariant for the kernels $P_\theta$ and $P_\mathbf{z}$ that respectively update $\theta$ and $\mathbf{z}$, and is therefore also invariant for the composite kernel $P=P_\theta P_\mathbf{z}$ \citep{Tierney.1994}. Moreover, since $P$ is strictly positive, the chain is $\pi$-irreducible and aperiodic. The chain is therefore positive Harris recurrent and thus ergodic.
\end{proof}

Hence, for any $\pi$-integrable function $g$, the ergodic theorem holds and the estimator
$\bar{g}_m := \frac{1}{m} \sum_{i=0}^{m-1} g\left(\theta^{(m)}, \mathbf{z}^{(m)}\right)$
is consistent for $E_\pi g$, for any initial distribution.

\section{Proof of Theorem \ref{theo:uni}}  \label{app:uni}

We start with a general theorem concerning DA-MCMC algorithms that alternate between Gibbs updates for the parameters and semi-independent M-H updates for the latent data; that is, the M-H proposal distribution does not dependent on the current configuration of the latent data.

\begin{theorem}  \label{theo:gen}
    Consider the Markov chain $\{(\theta_n, \mathbf{z}_n), n=1,2,\dots\}$ on $\chi = \chi_\theta \times \chi_\mathbf{z}$ corresponding to a DA-MCMC algorithm with target distribution $\pi$ and composite kernel $P = P_{\theta} P_{\mathbf{z}}$, where $P_{\theta}$ corresponds to a Gibbs sampler for $\theta$, and $P_{\mathbf{z}}$ to a semi-independent Metropolis-Hastings sampler for $\mathbf{z}$.
    If there exists positive functions $k_\theta$ and $k_r$ such that, for all $\mathbf{z} \in \chi_\mathbf{z}$,
    \begin{equation}
    \label{eq:kz}
        k_\theta(\theta) \le \pi(\theta | \mathbf{z}),
    \end{equation}
    and 
    \begin{equation}
    \label{eq:ktheta}
    k_r(\theta) \le \dfrac{q(\mathbf{z}|\theta)}{L(\theta; \mathbf{z})},
    \end{equation}
    where $q$ is the density of the Metropolis-Hastings proposal distribution and $L$ is the model likelihood,
    then the space $\chi$ is a \textit{small set}; as a result, the Markov chain is uniformly ergodic.
\end{theorem}

\begin{proof}
	By Proposition 2 in \cite{Tierney.1994}, it suffices to show that the state space $\chi = \chi_{\theta}\times\chi_{\mathbf{z}}$ is a \textit{small set} for the transition kernel $P$; that is, that there exists a probability measure $\nu$ on the $\sigma$-algebra $\sigma(\chi)$ such that
	\begin{equation}
		\label{eq:small}
		P^m(x,.) \ge \epsilon \nu(.), \quad \forall x\in \chi
	\end{equation}
	for a positive integer $m$ and constant $\epsilon > 0$. We show that inequality \eqref{eq:small} holds for $m=1$.

	Since $P = P_{\theta} P_{\mathbf{z}}$ is composite, with $P_{\theta}$ only updating $\theta$ and $P_{\mathbf{z}}$ only updating $\mathbf{z}$, we have
	\begin{equation}
	\label{eq:P}
	P((\theta_1, \mathbf{z}_1), (d\theta_2, d\mathbf{z}_2)) = P_\theta(\theta_1, d\theta_2 | \mathbf{z}_1) P_\mathbf{z}(\mathbf{z}_1, d\mathbf{z}_2 | \theta_2)
	\end{equation}
	where
	\begin{equation}
	\label{eq:Pth}
		P_\theta(\theta_1, d\theta_2 | \mathbf{z}_1)
		= \pi(d\theta_2 | \mathbf{z}_1)
		%= \pi(\beta_2 | \mathbf{z}_1) \pi(\lambda_2 | \mathbf{z}_1) d\beta_2 d\lambda_2
	\end{equation}
	corresponds to the transition kernel of a Gibbs sampler which does not depend on $\theta_1$ and
	\begin{align}
	\label{eq:Pz}
		P_\mathbf{z}(\mathbf{z}_1, d\mathbf{z}_2 | \theta_2)
		& = Q(d\mathbf{z}_2 | \theta_2) \alpha((\theta_2, \mathbf{z}_1),(\theta_2, \mathbf{z}_2)) + \delta_{\mathbf{z}_1}(d\mathbf{z}_2)\int(1-\alpha((\theta_2, \mathbf{z}_1),(\theta_2, \mathbf{z}_2))) Q(d\mathbf{z}_2| \theta_2) \nonumber \\
		& \ge Q(d\mathbf{z}_2| \theta_2) \alpha((\theta_2, \mathbf{z}_1),(\theta_2, \mathbf{z}_2)) \nonumber \\
		% & = q(\mathbf{z}_2| \theta_2) \min\left\lbrace 1, \dfrac{\pi(\theta_2, \mathbf{z}_2)q(\mathbf{z}_1| \theta_2)}{\pi(\theta_2, \mathbf{z}_1)q(\mathbf{z}_2| \theta_2)} \right\rbrace d\mathbf{z}_2 \nonumber \\
		& = q(\mathbf{z}_2| \theta_2) \min\left\lbrace 1, \dfrac{L(\theta_2; \mathbf{z}_2)q(\mathbf{z}_1| \theta_2)}{L(\theta_2; \mathbf{z}_1)q(\mathbf{z}_2| \theta_2)} \right\rbrace d\mathbf{z}_2
	\end{align}
	corresponds to a M-H transition kernel in which a new configuration of the latent data $\mathbf{z}_2$ is generated from the proposal kernel $Q$ conditionally on the current value of the parameters $\theta_2$, but independently of the current configuration $\mathbf{z}_1$.
		
	To show that $\chi$ is a small state, it thus suffices to find a positive function $k$ such that
	\begin{equation}
	\label{eq:min}
		k(\theta_2, \mathbf{z}_2) \le \pi(\theta_2 | \mathbf{z}_1) q(\mathbf{z}_2; \theta_2) \min\left\lbrace 1, \dfrac{L(\theta_2; \mathbf{z}_2)q(\mathbf{z}_1| \theta_2)}{L(\theta_2; \mathbf{z}_1)q(\mathbf{z}_2| \theta_2)} \right\rbrace, \quad \forall \mathbf{z}_1 \in \chi_\mathbf{z}.
	\end{equation}
	Indeed, suppose that we can find such function $k$, then for any set $A\in\sigma(\chi)$ and any $(\theta, \mathbf{z})\in\chi$ we have, by \eqref{eq:Pth}, \eqref{eq:Pz} and \eqref{eq:min},
	\begin{align*}
		P((\theta, \mathbf{z}), A)
		 \ge \int_A \pi(\theta' | \mathbf{z}) \min\left\lbrace 1, \dfrac{L(\theta_2; \mathbf{z}_2)q(\mathbf{z}_1| \theta_2)}{L(\theta_2; \mathbf{z}_1)q(\mathbf{z}_2| \theta_2)} \right\rbrace d(\theta', \mathbf{z}') 
		 \ge \int_A k(\theta', \mathbf{z}') d(\theta', \mathbf{z}') 
		 = \epsilon \nu(A)
	\end{align*}
	with $\epsilon = \int k(\theta, \mathbf{z}) d(\theta, \mathbf{z})$ a positive constant and $\nu(A) = \epsilon^{-1} \int_A k(\theta, \mathbf{z}) d(\theta, \mathbf{z})$ a probability measure.
		
	We can construct a positive function $k$ satisfying \eqref{eq:min} as follows. Note that inequality \eqref{eq:min} depends on $(\theta_1, \mathbf{z}_1)$ only through the full conditional distribution $\pi(\theta_2 | \mathbf{z}_1)$ and the ratio $\dfrac{q(\mathbf{z}_1| \theta_2)}{L(\theta_2; \mathbf{z}_1)}$, which are, by the assumptions \eqref{eq:ktheta} and \eqref{eq:kz} of the theorem, minorized by positive functions $k_\theta$ and $k_r$ respectively. The expression
	\begin{equation*}
	    k(\theta_2, \mathbf{z}_2) = k_\theta(\theta_2) q(\mathbf{z}_2; \theta_2) \min\left\lbrace 1, k_r(\theta_2) \dfrac{L(\theta_2; \mathbf{z}_2)}{q(\mathbf{z}_2|\theta_2)} \right\rbrace
	\end{equation*} 
	is therefore positive and satisfies inequality \eqref{eq:min}, which proves the theorem.
\end{proof}

To show the uniform ergocidity of our DA-MCMC algorithm, it therefore suffices to derive closed-form expressions for $k_r$ and $k_\theta$ in \eqref{eq:kz} and \eqref{eq:ktheta}. We start with with $k_r$.

\begin{proposition}  \label{pro:kr}
	The function 
	$k_r(\theta) = \dfrac{\prod_{k=1}^K \prod_{j\in \mathcal{I}_k} \exp\{-\beta n (t_k - t_{k-1})\}}{n^{n_I}}$
	is positive and satisfies inequality \eqref{eq:kz} for any distribution on the infection period.
\end{proposition}

\begin{proof}
	%\ram{(in the second inequality, minimize the whole numerator in terms of $I(.)$ instead of its each factors (=> use previous proof minimizing the proposal density); this will provide a better bound)}
	
	From \eqref{eq:cdl} and \eqref{eq:q}, we have
	\begin{align*}
		\dfrac{q(\mathbf{z}|\theta)}{L(\theta; \mathbf{z})} 
		& = \dfrac{\prod_{k=1}^K \prod_{j\in\mathcal{I}_k} \textsf{TrunExp}(z^I_j; \mu_{k}, t_{k-1}, t_{k})
		\prod_{l \in \mathcal{R}}f(z^R_l - z^I_l; \theta)
		\prod_{m \in \mathcal{R}^c} \bar{F}(T - z^I_m; \theta)}
		{\prod_{i \in \mathcal{I}} \beta I(z^I_i) \exp\left\lbrace - \int_{0}^{T}\beta S(t)I(t)dt \right\rbrace
		\prod_{l \in \mathcal{R}}f(z^R_l - z^I_l; \theta)
		\prod_{m \in \mathcal{R}^c} \bar{F}(T - z^I_m; \theta)} \\
		%& = && \dfrac{\prod_{k=1}^K \prod_{j\in \mathcal{I}_k} \textsf{TrunExp}(z^I_j; \mu_k, t_{k-1}, t_k)}
		%{\beta^{n_I} \prod_{i\in \mathcal{I}} I(z^I_i) \exp\left\lbrace - \beta \int_{0}^{T} S(t)I(t) dt \right\rbrace } \\
		%& && * \dfrac{\prod_{l=1}^n (1-p_l)^{\mathbf{1}\{z^R_l=\infty\}} \left(p_l \textsf{TrunWeib}(z^R_l - z^I_l; \lambda, a, 0, T - z^I_l)\right)^{\mathbf{1}\{z^R_l \le T\}}}
		%{\lambda^{n_R} a^{n_R} \prod_{m \in \mathcal{R}} (z^R_m - z^I_m)^{a-1}
		%\exp\left\lbrace - \lambda \left[ \sum_{m \in \mathcal{R}} (z^R_m - z^I_m)^a + \sum_{m \in \mathcal{R}^c} (T - z^I_m)^a\right] \right\rbrace} \\
		& = \dfrac{\prod_{k=1}^K \prod_{j\in \mathcal{I}_k} \textsf{TrunExp}(z^I_j; \mu_k, t_{k-1}, t_k)}
		{\beta^{n_I} \prod_{i\in \mathcal{I}} I(z^I_i) \exp\left\lbrace - \beta \int_{0}^{T} S(t)I(t) dt \right\rbrace } \\
		& = \dfrac{\prod_{k=1}^K \prod_{j\in \mathcal{I}_k} \dfrac{\beta I(t_{k-1}) \exp\{-\beta I(t_{k-1}) (z^I_j - t_{k-1})\}}{1-\exp\{-\beta I(t_{k-1})(t_k - t_{k-1})\}}}
		{\beta^{n_I} \prod_{i\in \mathcal{I}} I(z^I_i) \exp\left\lbrace - \beta \int_{0}^{T} S(t)I(t) dt \right\rbrace}\\
		%& = \dfrac{\prod_{k=1}^K \prod_{j\in \mathcal{I}_k} \dfrac{I(t_{k-1}) \exp\{-\beta I(t_{k-1}) (z^I_j - t_{k-1})\}}{1-\exp\{-\beta I(t_{k-1})(t_k - t_{k-1})\}}}
		%{\prod_{l\in \mathcal{I}} I(z^I_l) \exp\{-\beta \int_0^{T} S(t)I(t)dt\}}\\
		& \ge \dfrac{\beta^{n_I}\prod_{k=1}^K \prod_{j\in \mathcal{I}_k} I(t_{k-1}) \exp\{-\beta I(t_{k-1}) (z^I_j - t_{k-1})\}}		{\beta^{n_I} \prod_{i\in \mathcal{I}} I(z^I_i) \exp\left\lbrace - \beta \int_{0}^{T} S(t)I(t) dt \right\rbrace}\\
		& \ge \dfrac{\prod_{k=1}^K \prod_{j\in \mathcal{I}_k} \exp\{-\beta n (t_k - t_{k-1})\}}
		{n^{n_I}}.
	\end{align*}
	The first inequality holds because $1-\exp\{-\beta I(t_{k-1})(t_k - t_{k-1})\} \le 1$, and the second because $1 \le I(t) \le n$ for all $t$, $0\le\int_0^{T} S(t)I(t)dt$ and $z^I_j \le t_k$ when $j\in\mathcal{I}_k$.
\end{proof}

We now turn to $k_\theta$ in \eqref{eq:ktheta}. The following two lemmas are useful to find a closed-form expression for $k_\theta$  for our DA-MCMC.
% Gamma 1
\begin{lemma}  \label{lem:ga1}
	Let $\textsf{Ga}(x;a,b) = \dfrac{a^b}{\Gamma(a)}x^{a-1}\exp\{-ab\}$ denote the density of the gamma distribution with shape parameter $a$ and rate parameter $b$ evaluated at $x$. Then
	\begin{equation}
		\label{eq:ga1}
		\inf_{0\le \beta\le B} \textsf{Ga}(x;a,b+\beta) = 
		\begin{cases}
			\textsf{Ga}(x;a,b  ), & x <   x_a^* \\ \textsf{Ga}(x;a,b+B), & x \ge x_a^*
		\end{cases}
	\end{equation}	
	where $x_a^*=\frac{a}{B}\log\left( 1+\frac{B}{b}\right) $. Moreover,
	\begin{equation}
		\label{eq:ga2}
		\inf_{0\le \alpha\le A} \textsf{Ga}(x;a+\alpha,b) = 
		\begin{cases}
			\textsf{Ga}(x;a  ,b), & x >   x_b^* \\ \textsf{Ga}(x;a+A,b), & x \le x_b^*
		\end{cases}
	\end{equation}
	where $x_b^*=\frac{1}{b}\left[ \frac{\Gamma(a+A)}{\Gamma(a)}\right]^{1/A}$. 
\end{lemma}

\begin{proof}
	Equation \eqref{eq:ga1} is proven in \cite{Jones.2004}. For Equation \eqref{eq:ga2}, note that $x_b^*$ is the only positive solution to $\textsf{Ga}(x;a,b) = \textsf{Ga}(x;a+A,b)$. Now, for all $0<x\le x_b^*$ and all $0\le\alpha\le A$, we have
	\begin{align*}
		\frac{\textsf{Ga}(x;a+A,b)}{\textsf{Ga}(x;a+\alpha,b)}
		& = b^{A-\alpha}x^{A-\alpha} \frac{\Gamma(a+\alpha)}{\Gamma(a+A)} \\
		& \le b^{A-\alpha}\left(\frac{1}{b}\left[ \frac{\Gamma(a+A)}{\Gamma(a)}\right]^{1/A} \right)^{A-\alpha} \frac{\Gamma(a+\alpha)}{\Gamma(a+A)} \\
		& = \left[ \frac{\Gamma(a+A)}{\Gamma(a)}\right]^{(A-\alpha)/A} \frac{\Gamma(a+\alpha)}{\Gamma(a+A)} \\
		& = \left[
		\frac{\left(\frac{\Gamma(a+A)}{\Gamma(a)}\right)^{1/A}}{\left(\frac{\Gamma(a+A)}{\Gamma(a+\alpha)}\right)^{1/(A-\alpha)}}
		\right]^{A-\alpha} \\
		& = \left( \frac{\Gamma_{a,a+A}}{\Gamma_{a+\alpha, a+A}}\right)^{A-\alpha} \\
		& \le 1,
	\end{align*}
	where $\Gamma_{d,e} = \left( \frac{\Gamma(e)}{\Gamma(d)} \right)^{\frac{1}{e-d}}$ is a geometric mean,
	and where the last inequality holds because $\Gamma_{a,a+A}\le\Gamma_{a+\alpha, a+A}$.
	The case $x>x_b^*$ is analogous and is omitted for brevity.
\end{proof}

\begin{figure}
	\centering
	\subfloat[{$\alpha \in [0, 1]$, $\beta = 0$         }]{\includegraphics[width = .32\textwidth]{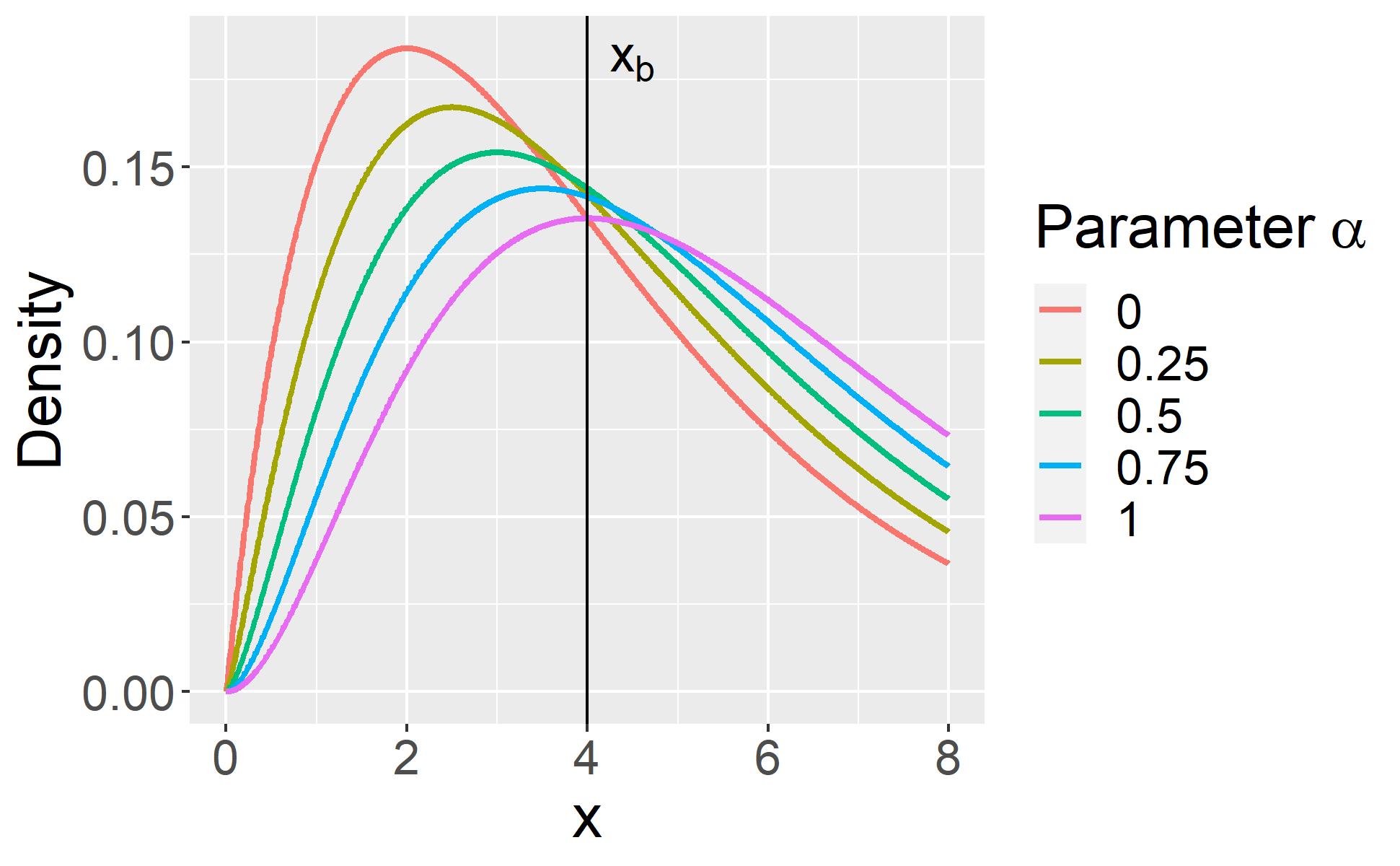} \label{fig:gam1a}}
	\subfloat[{$\alpha = 0$, $\beta \in [0, 1]$         }]{\includegraphics[width = .32\textwidth]{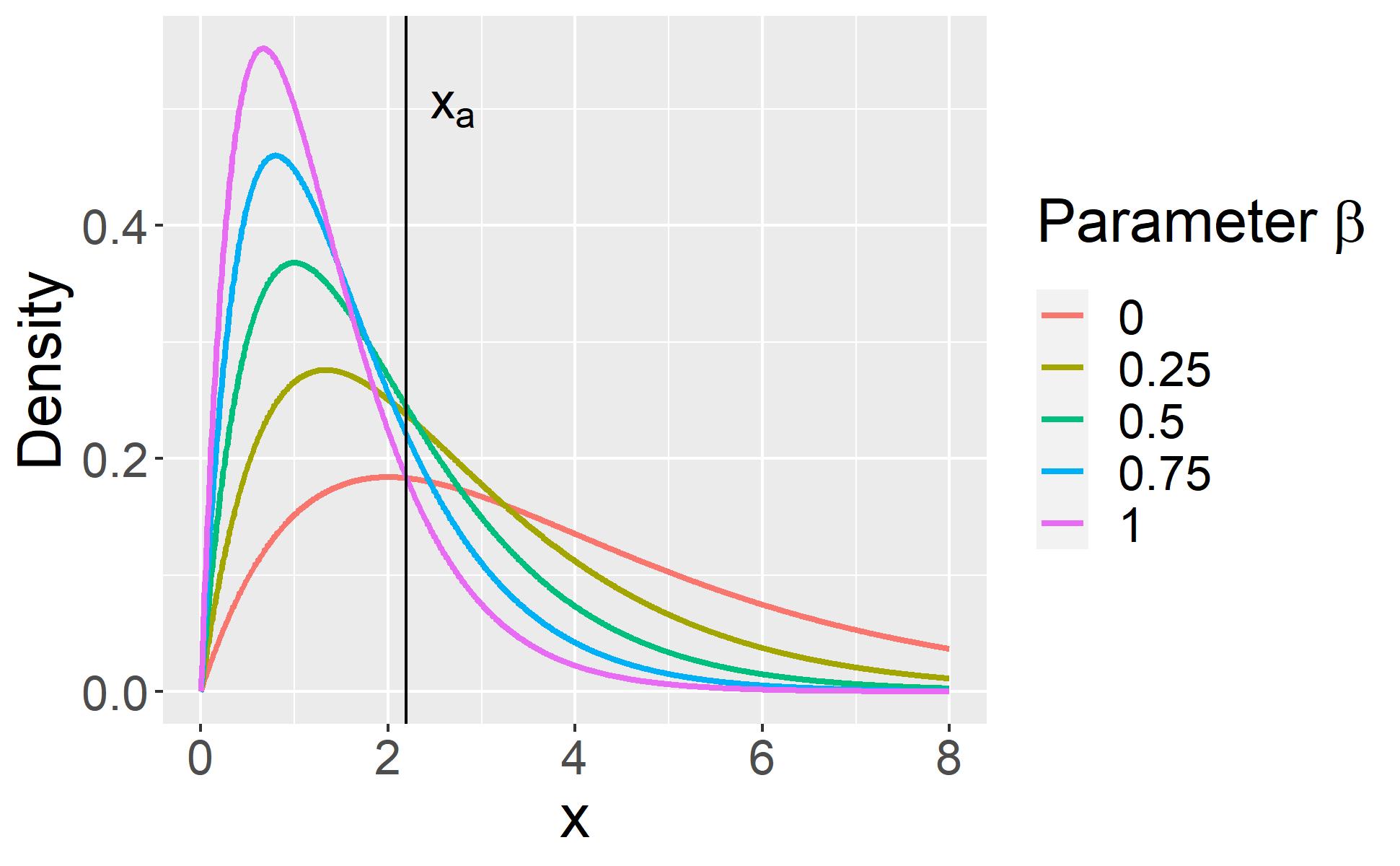} \label{fig:gam1b}}
	\subfloat[{$\alpha \in \{0,1\}$, $\beta \in \{0,1\}$}]{\includegraphics[width = .32\textwidth]{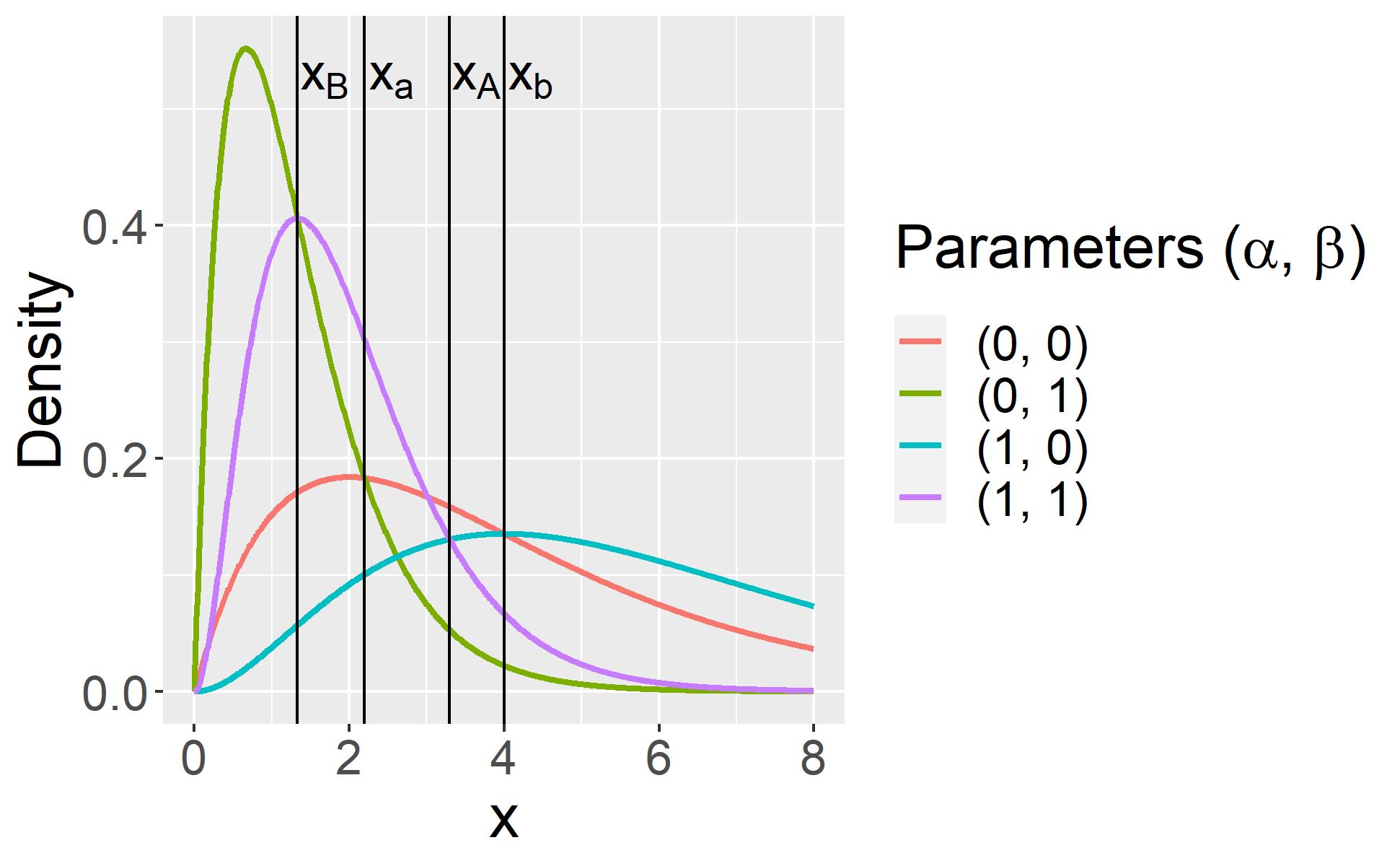} \label{fig:gam2}}
	
	\caption{Example of minorization of the gamma density $\textsf{Ga}(2+\alpha, 0.5+\beta)$.}
	\label{fig:gam}
\end{figure}

Figures \ref{fig:gam1a} and \ref{fig:gam1b} illustrate these two results. Lemma \ref{lem:ga1} can be used to obtain a closed form solution to the minimization of a gamma density of the form
\begin{equation}
	\label{eq:ga}
	\textsf{Ga}(x;a+\alpha,b+\beta) = \frac{(b+\beta)^{a+\alpha}}{\Gamma(a+\alpha)}x^{a+\alpha-1}\exp\{-x(b+\beta)\}, \quad 0\le\alpha\le A, 0\le\beta\le B.
\end{equation}
jointly over $(\alpha, \beta)$ for a given $x$. To this end, we establish the following technical lemma.
% Gamma 2 (joint)
\begin{lemma}  \label{lem:ga2}
	For $0 \le \alpha \le A$, $0 \le \beta \le B$ and a fix $x>0$, the density $\textsf{Ga}(x;a+\alpha,b+\beta)$ is minimized by $(\alpha, \beta) \in \{(A, 0), (0, B)\}$, the minimizing set of values depending on $x$. In particular,	
	$$
	\inf_{\begin{aligned}
			0\le \alpha\le A \\ 0\le \beta\le B
	\end{aligned}}\textsf{Ga}(x;a+\alpha,b+\beta) = 
	\begin{cases}
		\textsf{Ga}(x;a+A,b)                     , & x<x_a         \text{ or } x<x_{a+A}^* \vee x_b^*\\
		\textsf{Ga}(x;a,b+B)                     , & x_{a+A}^* < x \text{ or } x_a^* \wedge x_{b+B}^* < x\\
		\textsf{Ga}(x;a+A,b) \wedge \textsf{Ga}(x;a,b+B), & x_a^* \wedge x_b^*\le x < x_{a+A}^* \vee x_{b+B}^*\\
	\end{cases}
	$$
\end{lemma}

\begin{proof}
	Given $x>0$, \eqref{eq:ga1} shows that for a fixed $\alpha$, the gamma density \eqref{eq:ga} is minimized by $\beta \in \{0, B\}$. Similarly,  \eqref{eq:ga2} shows that for a fixed $\beta$, $\alpha \in \{0, A\}$ minimizes \eqref{eq:ga}. This implies that for a fixed $x>0$, \eqref{eq:ga} is minimized by $(\alpha, \beta) \in \{(0,0), (A, 0), (0, B), (A,B)\}$.
	
	A case-by-case analysis of the nine possibilities
	$$
	(\{x<x_a^*\}; \{x_a^*<x<x_{a+A}^*\}; \{x_{a+A}^*<x\} )
	\times(\{x<x_{b+B}^*\}; \{x_{b+B}^*<x<x_b^*\}; \{x_b^*<x\})
	$$
	is presented in Table \ref{tab:ga9} and shows that it is sufficient to consider $(\alpha, \beta) \in \{(A, 0), (0, B)\}$.
	
	\begin{table}
	\caption{\label{tab:ga9} Values of $(\alpha, \beta)$ that minimize $\textsf{Ga}(x;a+\alpha,b+\beta)$ for different values of $x$.}
	\centering
	\fbox{%
	\begin{tabular}{l|c c c}
		{ }& $x<x_a^*$  & $x_a^*\le x<x_{a+A}^*$ & $x_{a+A}^* < x$ \\
		\hline
		$x<x_b^*$              & $(a+A, b)$ & $(a+A, b)$             &                 \\
		$x_b^*\le x<x_{b+B}^*$ & $(a+A, b)$ &  ad hoc                & $(a, b+B)$      \\
		$x_{b+B}^* < x$        &            & $(a, b+B)$             & $(a, b+B)$ \\
		\hline
	\end{tabular}
    }
    \end{table}

	The two empty entries in Table \ref{tab:ga9} correspond to configurations that are impossible. Indeed, for all $a, A, b, B$ we must have $x_b^* \ge x_a^*$ since
	$$
	\frac{x_b^*}{x_a^*} = \dfrac{\left(\frac{\Gamma(a+A)}{\Gamma(a)}\right)^{1/A} b^{-1} }{a B^{-1} \log\left( 1+B/b\right)} = \dfrac{\left(\frac{\Gamma(a+A)}{\Gamma(a)}\right)^{1/A}}{a} \dfrac{B/b}{\log\left( 1+B/b\right)} = \dfrac{\Gamma_{a,a+A}}{a} \dfrac{B/b}{\log\left( 1+B/b\right)} \ge 1
	$$
	where the inequality holds since $a \le \Gamma_{a,a+A}$ and $y \ge \log\left( 1+y\right)$,
	and similarly $x_{a+A}^* > x_{b+B}^*$ since
	$$
	\frac{x_{b+B}^*}{x_{a+A}^*} = \dfrac{\left(\frac{\Gamma(a+A)}{\Gamma(a)}\right)^{1/A} (b+B)^{-1} }{(a+A) B^{-1} \log\left( 1+B/b\right)} = \dfrac{\Gamma_{a,a+A}}{a+A} \dfrac{B}{b+B} \dfrac{1}{\log\left(1+B/b\right)} \le 1
	$$
	where the inequality holds since $\Gamma_{a,a+A} \le a+A$ and $\dfrac{B}{b+B} < \dfrac{B}{b} \le \log\left(1+B/b\right)$.
	
	Finally, if $x_a^* \wedge x_{b+B}^*\le x < x_{a+A}^* \vee x_{b}^*$, corresponding to the middle entry of the center column in Table \ref{tab:ga9}, then one needs to directly check which set of values in $\{(0,0), (A, 0), (0, B), (A,B)\}$ minimizes \eqref{eq:ga}. In fact, it is sufficient to consider only $\{(A,0), (0, B)\}$ since
	$$
	\textsf{Ga}(x, a+A, b) \le \begin{cases}
	\textsf{Ga}(x;a,b), & x < x_b^* \\
	\textsf{Ga}(x;a+A,b+B), & x < x_{a+A}^* 
	\end{cases}
	$$
	and
	$$
	\textsf{Ga}(x, a, b+B) \le \begin{cases}
	\textsf{Ga}(x;a,b), & x > x_a^* \\
	\textsf{Ga}(x;a+A,b+B), & x > x_{b+B}^* 
	\end{cases}
	$$
\end{proof}

\begin{proposition}  \label{pro:ktheta_min}
	The function 
	\begin{align}
	k_\theta(\theta)
	= & \min\{\textsf{Ga}(\beta;a_{\beta}+n_I, b_{\beta}), \textsf{Ga}(\beta;a_{\beta}+n_I,b_{\beta}+n(n_I+I_0) T)\} \nonumber \\
	& * \min\{\textsf{Ga}(\lambda; a_\lambda+n_I+I_0,b_\lambda), \textsf{Ga}(\lambda; a_\lambda,b_\lambda+(n_I+I_0) (T)^a))\}
	\end{align}
	is positive and satisfies inequality \eqref{eq:ktheta}.
\end{proposition}

\begin{proof}
	Note that $\pi(\theta| \mathbf{z}) = \pi(\beta| \mathbf{z})\pi(\lambda| \mathbf{z})$. 
	We first minorize $\pi(\beta| \mathbf{z})$ using Lemma \ref{lem:ga1}. Writing $C=\int_{0}^{T} I(t)S(t)dt$, we have $C \in (0, n(n_I+I_0) T]$ since $1 \le I(.) \le n_I+I_0$ and $0\le S(.) \le n$. By \eqref{eq:posterior_beta}, we have
	\begin{align*}
		\pi(\beta| \mathbf{z}) 
		& = \textsf{Ga}\left( a_{\beta} + n_I , b_{\beta} + C \right) \\
		& \ge \inf_{\begin{aligned}
			0 \le C \le n(n_I+I_0) T
		\end{aligned}}   Ga\left(\beta; a_{\beta} + n_I, b_{\beta} + C\right) \\
		& = \min\{\textsf{Ga}(\beta;a_{\beta}+n_I, b_{\beta}), \textsf{Ga}(\beta;a_{\beta}+n_I,b_{\beta}+n(n_I+I_0) T)\}.
	\end{align*}
	Note that we do not need to minimize over the number of observed infections $n_I$ since this value is fixed by $\mathbf{Y}$.
	
	We similarly minorize $\pi(\lambda| \mathbf{z})$ using Lemma \ref{lem:ga2}. Writing $D=\sum_{k \in \mathcal{R}} (\tau^R_k - \tau^I_k)^a + \sum_{k \in \mathcal{R}^c} (T - \tau^I_k)^a$, we have $D \in (0, (n_I+I_0) (T)^a]$ since $|\mathcal{R}| + |\mathcal{R}^c| = |\mathcal{I}| = n_I + I_0$,  $(\tau^R_k - \tau^I_k)^a < (T)^a$ and $(T - \tau^I_k)^a < (T)^a$. Moreover, the observed number of removals $n_R$ is bounded above by the sum of the number of observed infection $n_I$ and the number of initially infectious individuals $I_0$. By \eqref{eq:posterior_lambda}, we have
	\begin{align*}
		\pi(\lambda| \mathbf{z}) 
		& = \textsf{Ga}\left(a_\lambda + n_R, b_\lambda + D \right)  \\
		& \ge \inf_{\begin{aligned}
			0 \le n_R &\le n_I + I_0, \\
			0 \le D   &\le (n_I+I_0) (T)^a
		\end{aligned}} Ga\left(\lambda; a_\lambda + n_R, b_\lambda + D \right) \\
		& = \min\{\textsf{Ga}(\lambda; a_\lambda+n_I+I_0,b_\lambda), \textsf{Ga}(\lambda; a_\lambda,b_\lambda+(n_I+I_0) (T)^a))\}.
	\end{align*}
\end{proof}

\end{document}